\documentclass[a4paper,12pt]{article}
\usepackage[left=2.2cm,right=2.2cm,top=2cm,bottom=2.5cm]{geometry}

\renewcommand{\arraystretch}{1.2}

\usepackage[pdftex]{graphicx}
\graphicspath{{plots/}}

\usepackage{lscape} 
\usepackage{array}

\usepackage{fix-cm}

\usepackage[makeroom]{cancel}
\usepackage[normalem]{ulem}
\usepackage{amsmath,amssymb}
\usepackage{slashed}
\usepackage{cite}
\usepackage{pdflscape}
\usepackage{multirow}
\usepackage[thinlines]{easytable}
\usepackage{url}
\usepackage[utf8]{inputenc}
\usepackage{longtable}
\usepackage{booktabs}

\usepackage{braket}
\usepackage{siunitx}
\usepackage{verbatim}
\usepackage[dvipsnames]{xcolor}
\usepackage{xspace}
\usepackage{hyperref}
\usepackage{caption}

\usepackage{tocloft}
\allowdisplaybreaks 

\newcommand{\MeV}{\,{\rm MeV}}
\newcommand{\GeV}{\,{\rm GeV}}
\newcommand{\refeq}[1]{Eq.~(\ref{eq:#1})}
\newcommand{\refeqs}[2]{Eqs.~(\ref{eq:#1})-(\ref{eq:#2})}

\newcommand{\refeqa}[2]{Eqs.~(\ref{eq:#1}) and (\ref{eq:#2})}

\newcommand{\refsec}[1]{Sec.~\ref{sec:#1}}

\newcommand{\refapp}[1]{App.~\ref{app:#1}}
\newcommand{\reftab}[1]{Tab.~\ref{tab:#1}}
\newcommand{\reffig}[1]{Fig.~\ref{fig:#1}}
\newcommand{\T}{\mathcal{T}}
\newcommand{\F}{\mathcal{F}}
\renewcommand{\H}{\mathcal{H}}

\newcommand{\cO}{\mathcal{O}}
\newcommand{\order}[1]{\mathcal{O}\left(#1\right)}

\renewcommand{\Im}{\text{Im}\,}
\newcommand{\had}{\text{had}}
\newcommand{\OPE}{\text{OPE}}

\newcommand{\proj}{
        \frac{\slashed{n}_-\slashed{n}_+}{4}
}
\newcommand{\projb}{
        \frac{\slashed{n}_+\slashed{n}_-}{4}
}

\newcounter{TODO}

\newcommand{\ara}[1]{
    #1
    }

\makeatletter
    
    \def\tf{\@ifstar\@@tf\@tf}
    \newcommand{\@tf}[1]{{\color{purple}{[\textbf{TF:} #1]}}}
    \newcommand{\@@tf}[1]{{\color{purple}{#1}}}

    \def\ng{\@ifstar\@@ng\@ng}
    \newcommand{\@ng}[1]{{\color{ForestGreen}{[\textbf{NG:} #1]}}}
    \newcommand{\@@ng}[1]{{\color{ForestGreen}{#1}}}
\makeatother

\begin{document}

\begin{center}
\bf
\fontsize{18.1}{24}\selectfont 
\boldmath
Non-factorisable Contributions\\  of Strong-Penguin Operators in $\Lambda_b \to \Lambda \ell^+\ell^-$ Decays

\end{center}

\vspace{-0.2cm}

\begin{center}
\renewcommand{\thefootnote}{\fnsymbol{footnote}}
{Thorsten Feldmann$^a$, Nico Gubernari$^{a,b}$}
\renewcommand{\thefootnote}{\arabic{footnote}}
\setcounter{footnote}{0}

\vspace*{.3cm}

\centerline{${}^a$\it Theoretische Physik 1, Center for Particle Physics Siegen, Universit\"at Siegen, }
\centerline{\it Walter-Flex-Stra\ss{}e 3, 57068  Siegen,  Germany}
\vspace{1.3mm}
\centerline{${}^b$\it DAMTP, University of Cambridge,}
\centerline{\it  Wilberforce Road, Cambridge, CB3 0WA, United Kingdom}
\vspace{1.3mm}

\vspace*{-.2cm}

\end{center}

\begin{center} 

\textit{E-mail: {\sf thorsten.feldmann@uni-siegen.de,  nicogubernari@gmail.com}}

\vspace*{.2cm}

{\sf Preprint: \tt{SI-HEP-2023-36, P3H-23-104}}

\end{center}

\vspace{-.4cm}

\begin{abstract}
\noindent
We investigate for the first time a certain class of non-factorisable contributions of the four-quark operators $\cO_{3-6}$ in the weak effective Hamiltonian to the $\Lambda_b \to \Lambda \ell^+\ell^-$ decay amplitude. 
We focus on the case where a virtual photon is radiated from one of the light constituents of the $\Lambda_b$ baryon, in the kinematic situation of large hadronic recoil with an energetic $\Lambda$ baryon in the final state.
The effect on the suitably defined ``non-local form factors'' is calculated using the light-cone sum rule approach for a correlator with an interpolating current for the light $\Lambda$ baryon. We find that this approach requires the introduction of new soft functions that generalise the standard light-cone distribution amplitudes (LCDAs) for the heavy $\Lambda_b$ baryon. We give a heuristic discussion of their properties and a model that relates them to the standard LCDAs.
Within this framework, we provide numerical results for the size of the non-local form factors considered.
\end{abstract}

\vspace{-.4cm}


\setcounter{tocdepth}{2}
\tableofcontents

\renewcommand{\theequation}{\arabic{section}.\arabic{equation}}

\section{Introduction}
\setcounter{equation}{0}

Rare $b$-quark decays in general, and rare $b \to s \ell^+\ell^-$ transitions in particular, have received much attention in the past. 
From a phenomenological point of view, these decays provide a large number of complementary observables that allow us to explore the flavour sector of the Standard Model (SM) and its possible new-physics extensions (for comprehensive reviews, see e.g. Refs.~\cite{LHCb:2012myk,Belle-II:2018jsg}
and references therein). From a theoretical point of view, the decays of a heavy quark into light degrees of freedom provide a valuable playground to develop and refine calculation methods to address the factorisation of hadronic bound-state effects from short-distance QCD corrections. In particular, one can establish QCD factorisation theorems (see e.g. Refs.~\cite{Beneke:1999br,Beneke:2000ry,Beneke:2001ev} for the pioneering works) for decay amplitudes of exclusive decays with large energy transfer to one or more light hadrons in the final state. Here, the bound-state effects are contained in hadronic transition form factors of local decay currents and light-cone distribution amplitudes (LCDAs) for light and heavy hadrons. 
Alternatively, it is possible to replace one or the other hadron by suitably chosen interpolating currents and relate the exclusive decay amplitude to the corresponding correlators by dispersion relations, namely the light-cone sum rule (LCSR) method (for a recent review, see Ref.~\cite{Khodjamirian:2023wol}).
In both cases, the factorisation of ``soft'' degrees of freedom in the $b$-hadron and the energetic degrees of freedom in the final state can be formally achieved by matching onto a soft-collinear effective theory (SCET), see e.g. Refs.~\cite{Bauer:2000yr,Bauer:2001yt,Beneke:2002ph} for early applications in $b$-decays.

In the past, much of the phenomenological study of rare semileptonic $b \to s\ell^+\ell^-$ transitions has been devoted to exclusive $B \to K\ell^+\ell^-$ and $B \to K^*\ell^+\ell^-$ decays. In particular, a number of ``flavour anomalies'' (i.e.\ deviations between experimental measurements and theoretical expectations within the SM) may be due to physics beyond the SM, for recent reviews see Refs.~\cite{Albrecht:2021tul,Capdevila:2023yhq}.
However, a correct interpretation of these anomalies requires the control of systematic experimental effects as well as hadronic uncertainties in the theoretical predictions. For this reason, independent cross-checks of $b \to s\ell^+\ell^-$ transitions with complementary sensitivity to the different short-distance coefficients associated with the low-energy effective operators are desirable.

\begin{figure}[t!bp]
	\centering
	\includegraphics[width=0.95\textwidth]{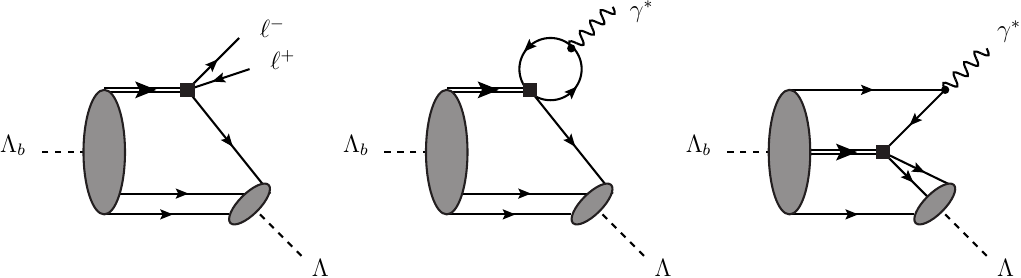}
	\caption{Illustration of different decay topologies for $\Lambda_b \to \Lambda \ell^+\ell^-$: Left: local form-factor contribution for semi-leptonic operators $\cO_9$ or $\cO_{10}$. Center: Non-local contribution from virtual photon radiation off a charm (or light) quark loop. Right: example for a non-local contribution of the four-quark operators $\cO_{3-6}$ in $\Lambda_b \to \Lambda \ell^+\ell^-$ decays. From the analogy to the corresponding terminology in $B \to K^{(*)}\ell^+\ell^-$ decays, this is referred to as an ``annihilation topology''. -- 
     In each diagram, operators from the weak effective Hamiltonian are indicated by a black square, the $b$-quark is indicated by a double line, and the virtual photon subsequently dissociates into $\ell^+\ell^-$ (not shown). 
    }
	\label{fig:process1}
\end{figure}

For instance, the angular analysis of baryonic transitions such as $\Lambda_b \to \Lambda(\to p\pi) \ell^+\ell^-$ \cite{LHCb:2018jna}
provides a number of independent observables \cite{Gutsche:2013pp,Boer:2014kda}  that can be combined with data on mesonic decays in a global fit, see e.g. Refs.~\cite{Blake:2019guk,Alguero:2021anc}.
The most important hadronic input functions in these analyses are the $\Lambda_b \to \Lambda$ (local) transition  form factors, which appear after factorising the (local) hadronic and leptonic currents in the semi-leptonic and electromagnetic penguin operators in the weak effective Hamiltonian, see the illustration in the left panel of Fig.~\ref{fig:process1}. Quantitative results for the form factors can be obtained from lattice-QCD studies \cite{Detmold:2016pkz,Meinel:2023wyg} for small recoil energy (large invariant lepton mass squared $q^2$) and LCSRs for large recoil energy,
see e.g. Refs.~\cite{Wang:2009hra,Aliev:2010uy,Feldmann:2011xf,Wang:2015ndk}.
Furthermore, one can exploit unitarity constraints in the form of a ``$z$-expansion'', which allows one to interpolate between small and large values of $q^2$ in a controlled way, see e.g. Ref.~\cite{Blake:2022vfl}.
Finally, in the heavy-quark limit for the $b$-quark, the number of independent form factors is reduced from ten to two at small recoil, and to only one at large recoil \cite{Mannel:1997xy,Mannel:2011xg,Feldmann:2011xf}.
Note that the numerically dominant part of the form-factor values at large recoil is due to the so-called ``soft-overlap'' mechanism, where the energy transfer to the final state cannot be described by a finite number of virtual gluon exchanges (despite the fact that the latter mechanism is parametrically of leading power in QCD factorisation, see the discussion in Ref.~\cite{Wang:2011uv}).

Besides the local form-factor terms, time-ordered products of the hadronic operators in the weak effective Hamiltonian and the electromagnetic QED interaction also appear, leading to generalised hadronic matrix elements, often called ``non-local form factors''.
An important example in $b \to s \ell^+\ell^-$ transitions is the so-called 
``charm-loop effect'', where a $c\bar c$ pair is produced by a four-quark operator in the weak effective Hamiltonian and then annihilated by a virtual photon that subsequently decays into the charged lepton pair, see the illustration in the middle panel of Fig.~\ref{fig:process1}. This effect restricts the perturbative treatment within QCD factorisation to values of $q^2$ well below the onset of charmonium resonances.\footnote{
    By a similar reasoning, the considered values of $q^2$ should be taken \emph{above} the light vector resonances $\rho,\omega,\ldots$
    An attempt to model the low-$q^2$ spectrum associated to annihilation topologies in rare $D$-meson decays within the QCD factorization approach has been discussed in Ref.~\cite{Feldmann:2017izn}.
} Nevertheless, phenomenological information about the lowest-lying  $c\bar c$ resonances ($J/\psi$ and $\psi'$) can be implemented in a dispersive approach, which allows to extrapolate the perturbative results at small (or even negative) values to higher values of $q^2$. This has been worked out in some detail for mesonic transitions, see e.g. Refs.~\cite{Bobeth:2017vxj,Gubernari:2020eft,Gubernari:2022hxn}.
It is important to stress that an adequate knowledge of the size of the non-local form factors in rare $b$-hadron decays is essential for a reliable estimate of theoretical uncertainties.
Only the combination of accurate theoretical predictions and experimental measurements
of rare flavour observables allows us to constrain the size of physics beyond the SM in these decays or even establish deviations from SM predictions.

In this work, we are interested in decay topologies like the one shown on the right panel of \reffig{process1}, for which there is no estimate to date. Here the four-quark operators $\cO_{3-6}$ connect two pairs of quarks in the initial and final baryon, with a virtual photon emitted from one of these quarks dissociating into $\ell^+\ell^-$. 
This is analogous to the annihilation topologies in rare $B \to K^{(*)}\ell^+\ell^-$ decays, and hence we refer to this contribution as ``annihilation topologies''.
From the example shown, it is immediately clear that the third valence quark (plus additional gluons or $q\bar q$ pairs) does not necessarily participate in the hard-scattering process. In the context of QCD factorisation \cite{Beneke:2001ev} this would correspond to kinematic endpoint configurations which prevent a complete factorisation of the decay amplitude into hadronic LCDAs and short-distance kernels. In the following, we therefore consider the ``annihilation topologies'' in the framework of LCSRs with $b$-hadron LCDAs, following the procedures outlined in Refs.~\cite{DeFazio:2005dx,Khodjamirian:2006st} for mesonic form factors and Refs.~\cite{Feldmann:2011xf,Wang:2015ndk} for baryonic form factors.

The outline of this article is as follows. 
In \refsec{setup} we give our definitions of local and non-local form factors for the relevant operators in the weak effective Hamiltonian. We also introduce our kinematic notations and conventions that we use in the sum-rule calculation.
In \refsec{calc} we introduce the correlator from which we deduce the contribution of the annihilation topologies by comparing the perturbative 
calculation in leading-order QCD with the hadronic representation in terms of non-local form factors.
We find that the dominant effects are associated with the case where a virtual photon is emitted from one of the light-quarks in the $\Lambda_b$ baryon. In this situation we find that the soft hadronic function describing the bound-state properties of the $\Lambda_b$ baryon is given in terms of a tri-local operator where the two light-quark fields are separated along two \emph{different} light-cone directions. We derive the general momentum-space projector for this case in terms of generalised LCDAs. 
Sec.~\ref{sec:num} is devoted to the numerical analysis of the sum rule with a careful assessment of parametric uncertainties from various sources. 
In comparison with the corresponding contribution of the local form-factor terms in the SM, we observe that the leading effect of the annihilation topologies is of similar size as in the mesonic counterpart
for $B \to K \ell^+\ell^-$ or $B \to K_\parallel^*\ell^+\ell^-$.
In fact, we find $\order{1\%}$ effects at the amplitude level for transverse $\gamma^*$ polarisation, whereas the effect for longitudinal $\gamma^*$ polarisation turns out to be negligible. 
After our conclusions in Sec.~\ref{sec:conc}, we provide some additional details about our modelling of the generalised LCDAs and some of the calculation steps in the sum-rule calculation in Apps.~\ref{app:LCDAmodel} and \ref{app:match}.

\section{Theoretical framework}
\setcounter{equation}{0}
\label{sec:setup}

\subsection{Definition of local form factors}

In $\Lambda_b \to \Lambda \ell^+\ell^-$ decays, the naively factorising contributions from the operators 
\begin{align}
	\cO_9 &= 
	\frac{\alpha_e}{4\pi}
	\left(\bar s \gamma^\mu P_L b \right)
	\left(\bar \ell \gamma_\mu \ell \right) \,,
	\qquad 
	\cO_{10} = 
	\frac{\alpha_e}{4\pi}
	\left(\bar s \gamma^\mu P_L b \right)
	\left(\bar \ell \gamma_\mu \gamma_5 \ell \right) \,,
\end{align}
in the weak effective Hamiltonian require the knowledge of the local transition form factors for vector and axial-vector $b \to s$ currents.
Our conventions for these form factors in the helicity basis
 follow the definition in Ref.~\cite{Feldmann:2011xf}. For the vector form factor we use
\begin{align}
\F_\mu \equiv	\langle \Lambda(p')| \bar s \gamma_\mu b |\Lambda_b(p)\rangle 
	& = 
	\bar u_\Lambda(p') \left\{ 
	f_0(q^2) \, (M_{\Lambda_b} - m_\Lambda) \, \frac{q_\mu}{q^2}
	\right. 
	\nonumber\\*
	&  \qquad \left. {} + f_+(q^2) \, \frac{M_{\Lambda_b}+m_\Lambda}{s_+} 
	\left( p_\mu + p'_\mu - \frac{q_\mu}{q^2} \, (M_{\Lambda_b}^2-m_\Lambda^2) \right) \right. 
	\nonumber\\*
	& \qquad \left. {} + f_\perp(q^2) \left(\gamma_\mu - \frac{2m_\Lambda}{s_+} \, p_\mu - 
	\frac{2m_{\Lambda_b}}{s_+} \, p'_\mu\right) 	
	\right\} u_{\Lambda_b}(p) \,,
\end{align}
where
\begin{equation}
	s_\pm = (M_{\Lambda_b} \pm m_\Lambda)^2 - q^2 \,.
\end{equation}
An analogous definition holds for the form factors of axial-vector currents,
\begin{align}
\F_{\mu 5} \equiv	\langle \Lambda(p')| \bar s \gamma_\mu \gamma_5 b |\Lambda_b(p)\rangle 
	& = 
	-\bar u_\Lambda(p') \gamma_5 \left\{ 
	g_0(q^2) \, (M_{\Lambda_b} + m_\Lambda) \, \frac{q_\mu}{q^2}
	\right. 
	\cr 
	&  \qquad \left. {} + g_+(q^2) \, \frac{M_{\Lambda_b}-m_\Lambda}{s_-} 
	\left( p_\mu + p'_\mu - \frac{q_\mu}{q^2} \, (M_{\Lambda_b}^2-m_\Lambda^2) \right) \right. 
	\cr 
	& \qquad \left. {} + g_\perp(q^2) \left(\gamma_\mu + \frac{2m_\Lambda}{s_-} \, p_\mu - 
	\frac{2m_{\Lambda_b}}{s_-} \, p'_\mu\right) 	
	\right\}  \, u_{\Lambda_b}(p) \,,
\end{align}
Throughout this work the spin arguments for the fermion states and Dirac spinors are usually not explicitly shown for simplicity, i.e. $u_{\Lambda_b}(p,s)\equiv u_{\Lambda_b}(p)$.
The projections on the vector form factors read
\begin{align}
	f_0(q^2) \left[ \bar u_\Lambda(p') \, u_{\Lambda_b}(p) \right] & = 
	    \frac{q^\mu \F_\mu}{M_{\Lambda_b} - m_\Lambda} \,,
	    \cr 
	f_+(q^2) \left[\bar u_\Lambda(p') \, u_{\Lambda_b}(p) \right] 
	& = 
	- \frac{2 q^2}{s_-} \, \frac{p^\mu \F_\mu}{M_{\Lambda_b} + m_{\Lambda}} \,, 
	\cr 
	f_\perp(q^2) \left[\bar u_\Lambda(p') \, \gamma^\nu_\perp \, u_{\Lambda_b}(p) \right] 
	& = g^{\nu\mu}_\perp \F_\mu \,,
\end{align}
and similarly for the axial-vector form factors,
where 
$$
 g^{\mu\nu}_\perp = g^{\mu\nu} + \frac{4 m_{\Lambda}^2}{s_+ s_-} \, p^\mu p^\nu + \frac{4 M_{\Lambda_b}^2}{s_+ s_-} \, p'{}^\mu p'{}^\nu 
 - \left( \frac{1}{s_+} + \frac{1}{s_-} \right) \left( p^\mu p'{}^\nu + p'{}^\mu p^\nu \right)
$$
is the metric tensor in the plane transverse to the momentum vectors $p$ and $p'$.

\subsection{Definition of non-local form factors}

In this work, we are interested in hadronic matrix elements of the time-ordered product of the strong-penguin operators in the weak effective Hamiltonian
\begin{equation}
\begin{aligned}
    \cO_3 & =
        \left(\bar{s} \gamma^\nu P_L b  \right)
        \sum_{q}
        \left(\bar{q} \gamma_\nu P_L\, q  \right)
    \,,&&
    & \cO_4  =
        \left(\bar{s}^j \gamma^\nu P_L b^i  \right)
        \sum_{q}
        \left(\bar{q}^i \gamma_\nu P_L\, q^j  \right)
    \,,\\
    \cO_5 & =
        \left(\bar{s} \gamma^\nu P_L b  \right)
        \sum_{q}
        \left(\bar{q} \gamma_\nu P_R\, q  \right)
    \,,&&
    & \cO_6  =
        \left(\bar{s}^j \gamma^\nu P_L b^i  \right)
        \sum_{q}
        \left(\bar{q}^i \gamma_\nu P_R\, q^j  \right)
    \,,
\end{aligned}
\label{eq:O36}
\end{equation}
with the electromagnetic current
\begin{align}
    \label{eq:jem}
	J_\mu^{\rm em}(x) = \sum_{q} Q_q \, \bar q(x) \, \gamma_\mu \, q(x) \,,
\end{align}
where $Q_q$ is the $q$ quark charge.
These matrix elements can be decomposed in the same way as the local form factors:
\begin{align}
	\H_\mu^{(i)}
	& =
	i\int d^4x\, e^{i q\cdot x}
	\bra{\Lambda(p')}
	\T\! \left\{
	\cO_{i}(0)
	J_\mu^{\rm em}(x)
	\right\}
	\ket{\Lambda_b(p)}  
	\cr  
	& \equiv M_{\Lambda_b}^2 \, \bar u_{\Lambda}(p') 
	\Bigg\{
	\left.
	\H_+^{(i)}(q^2) \, \frac{M_{\Lambda_b}+m_{\Lambda}}{s_+} \,
	\left(p_\mu + p_\mu' - \frac{q_\mu}{q^2} \, (M_{\Lambda_b}^2-m_{\Lambda}^2) \right)
	\right.
	\cr 
	& 
	\qquad \qquad  {} + \H_\perp^{(i)}(q^2) \left( \gamma_\mu - \frac{2 m_\Lambda}{s_+} \, p_\mu - \frac{2  M_{\Lambda_b}}{s_+} \, p '_\mu
	\right)
	\Bigg\} u_{\Lambda_b}(p) 
 \cr 
 & \phantom{\equiv} {} 
 +
 M_{\Lambda_b}^2 \, \bar u_{\Lambda}(p')  \gamma_5
	\Bigg\{
	\left.
	\H_{+5}^{(i)}(q^2) \, \frac{M_{\Lambda_b}-m_{\Lambda}}{s_-} \,
	\left(p_\mu + p_\mu' - \frac{q_\mu}{q^2} \, (M_{\Lambda_b}^2-m_{\Lambda}^2) \right)
	\right.
	\cr 
	& 
	\qquad \qquad  {} + \H_{\perp 5}^{(i)}(q^2) \left( \gamma_\mu + \frac{2 m_\Lambda}{s_-} \, p_\mu - \frac{2  M_{\Lambda_b}}{s_-} \, p '_\mu
	\right)
	\Bigg\}  \, u_{\Lambda_b}(p)
 \,.
	\label{eq:had-ffs}
\end{align}
Notice that due to the conservation of the electromagnetic current, Lorentz structures proportional to $q^\mu$ do not appear on the r.h.s of this equation.
In the following, we use the term ``non-local form factors'' for the generalised objects $\H_{+(5)}^{(i)}$ and $\H_{\perp(5)}^{(i)}$.
These non-local form factors  can be isolated by contracting \refeq{had-ffs} with $p^\mu$ and $g^{\mu\nu}_\perp$, respectively.
These projections read
\begin{equation}
\begin{aligned}
    p^\mu \H_\mu^{(i)}
    & = 
    -\frac{M_{\Lambda_b}^2}{2 q^2}\,
    u_\Lambda(p')
    \left[
        s_- (M_{\Lambda_b} + m_{\Lambda})
        \H_+^{(i)}(q^2)  
        - s_+ (M_{\Lambda_b} - m_{\Lambda})
        \H_{+5}(q^2) \gamma_5
    \right] u_{\Lambda_b}(p)
    \,, \\
    g^{\mu\nu}_\perp
    \H_\mu^{(i)} 
    & =     
    M_{\Lambda_b}^2
    u_\Lambda(p')
    \left[
    \H_\perp^{(i)}(q^2) 
     \, \gamma^\nu_\perp 
    -  \H_{\perp5}^{(i)}(q^2) 
    \gamma^\nu_\perp  
    \gamma_5 
    \right] u_{\Lambda_b}(p)
    \,.
 \label{eq:Hperp}
\end{aligned}    
\end{equation}
With these conventions, the non-local contributions to the $\Lambda_b \to \Lambda \ell^+\ell^-$ decay amplitude can be rewritten in terms of a $q^2$ dependent shift of $C_9$.
Hence they can be accounted for using the following replacements:
\begin{align}
    \label{eq:deltaC9a}
    C_9 \, f_+(q^2)  & \to C_9 \, f_+(q^2) - 16\pi^2 \,  \frac{2M_{\Lambda_b}^2}{q^2} \, \sum_{i} \, C_i \, \H_+^{(i)}(q^2)
    \equiv f_+(q^2) \left(C_9 + \Delta C_{9,+}(q^2)\right)
    \,,  
    \\ 
    \label{eq:deltaC9b}
    C_9 \, f_\perp(q^2)  & \to C_9 \, f_\perp(q^2) - 16\pi^2 \,  \frac{2M_{\Lambda_b}^2}{q^2} \, \sum_{i} \, C_i \, \H_\perp^{(i)}(q^2)
    \equiv f_\perp(q^2) \left(C_9 + \Delta C_{9,\perp}(q^2)\right)
    \,,
    \\ 
    \label{eq:deltaC9c}
    C_9 \, g_+(q^2)  & \to C_9 \, g_+(q^2) - 16\pi^2 \,  \frac{2M_{\Lambda_b}^2}{q^2} \, \sum_{i} \, C_i \, \H_{+5}^{(i)}(q^2)
    \equiv g_+(q^2) \left(C_9 + \Delta C_{9,+5}(q^2)\right)
    \,,  
    \\
    \label{eq:deltaC9d}
    C_9 \, g_\perp(q^2)  & \to C_9 \, g_\perp(q^2) - 16\pi^2 \,  \frac{2M_{\Lambda_b}^2}{q^2} \, \sum_{i} \, C_i \, \H_{\perp5}^{(i)}(q^2)
    \equiv g_\perp(q^2) \left(C_9 + \Delta C_{9,\perp 5}(q^2)\right)
    \,.
\end{align}
Similar relations are known for the mesonic counterpart $B \to K^{(*)}\ell^+\ell^-$, see, for instance, Refs.~\cite{Beneke:2001at,Bobeth:2017vxj}.

\subsection{Light-cone vectors and power counting}
\label{sec:pow}

It is convenient to introduce the following light-cone vectors
\begin{align}
	n_+^\mu &\equiv \frac{\left( \sqrt{s_+}+\sqrt{s_-} \right)^2}{2 M_{\Lambda_b} \, \sqrt{s_+ s_-}} \, p^\mu 
	- \frac{2 M_{\Lambda_b}}{\sqrt{s_+ s_-}} \, p'{}^\mu \,,
		\nonumber \\[0.2em]  
	n_-^\mu & \equiv  - \frac{\left( \sqrt{s_+}-\sqrt{s_-} \right)^2}{2 M_{\Lambda_b} \, \sqrt{s_+ s_-}} \, p^\mu 
	+ \frac{2 M_{\Lambda_b}}{\sqrt{s_+ s_-}} \, p'{}^\mu \,,	
\end{align}
such that 
\begin{align}
	n_+^2=n_-^2 = 0 \,, \qquad 
	n_+ \cdot n_- = 2 \,, \qquad 
	p^\mu = M_{\Lambda_b} \, \frac{n_+^\mu + n_-^\mu}{2}
	\equiv M_{\Lambda_b} \, v^\mu \,,
\end{align}
with $v^\mu$ being the four-velocity of the $\Lambda_b$ baryon,
and the transverse metric can simply be written as
\begin{align}
    \label{eq:gperp}
	g^{\mu\nu}_\perp = g^{\mu\nu} - \frac{n_+^\mu n_-^\nu+n_-^\mu n_+^\nu}{2} \,. 
\end{align}
One can decompose any momentum vector in light-cone coordinates and consider the 
scaling of the individual momentum projections with a small expansion parameter $\lambda \ll 1$.
In our case, we take
\begin{align}
	\lambda^2 \sim &\frac{\Lambda_{\rm QCD}}{m_b} \ll 1 \,,
\end{align}
with $m_b \sim M_{\Lambda_b}$ being the mass of the heavy $b$-quark.
To identify the power-counting of the various momenta, we use the short-hand notation
\begin{eqnarray}
	k^\mu &:& \left\{ ( n_+ \cdot k) , \ k_\perp , (n_- \cdot k) \right\} \sim (\lambda^a,\lambda^b,\lambda^c) \,,
\end{eqnarray}
where the powers of $\lambda$ indicate the momentum scaling in units of $m_b$.
For the $b$-quark, which is treated as a quasi-static colour source in the framework of heavy-quark effective theory (HQET), we thus have
\begin{eqnarray}
	\mbox{HQET $b$-quark} &:& \quad  p_b^\mu = m_b \, v^\mu  + \Delta k^\mu \,, \qquad \text{with} \quad 
	\Delta k^\mu \sim (\lambda^2,\lambda^2,\lambda^2) \,,
\end{eqnarray}
and $\Delta k^\mu$ is referred to as a \emph{soft} residual momentum.
Similarly,
the light quarks and gluons in the $\Lambda_b$ bound state have soft momentum scaling:
\begin{eqnarray}
	\mbox{soft momenta in $\Lambda_b$ baryon} &:& \quad k_s^\mu \sim (\lambda^2,\lambda^2,\lambda^2) \,,
\end{eqnarray}
with virtualities $k_s^2 \sim \lambda^4$.
In this work, we concentrate on the large-recoil region, where --- in the rest frame of $\Lambda_b$ --- the energy of the hadronic final state is of the order $M_{\Lambda_b}/2$.
The light constituents of the $\Lambda$ have \emph{collinear} momenta, scaling as
\begin{eqnarray}
	\mbox{collinear momenta in $\Lambda$ baryon} &:& \quad k_c^\mu \sim (1,\lambda^2,\lambda^4) \,.
\end{eqnarray}
with virtualities $k_c^2 \sim \lambda^4$.
Interactions between field modes with soft and collinear momenta are induced by \emph{hard-collinear} momenta,
\begin{eqnarray}
	\mbox{internal hard-collinear modes} &:& \quad k_{\rm hc}^\mu \sim (1,\lambda,\lambda^2)
\end{eqnarray}
with virtualities $k_{\rm hc}^2 \sim \lambda^2$. Here, the scaling $k_{\rm hc}^\perp \sim \lambda$ refers to hard-collinear modes in loops (while tree-level interactions would have $k_{\rm hc}^\perp \sim \lambda^2$).
Finally, the momentum transfer to the lepton pair is given by
\begin{eqnarray}
	q^\mu = (p - p')^\mu \sim (\lambda^2,0,1) \,, \qquad \mbox{with $q^2 \sim \lambda^2$ \, (anti-hard-collinear)}
\end{eqnarray}
and is referred to as \emph{anti-hard-collinear} (this excludes the kinematic limit $q^2 \to 0$).

\section{Calculation of the ``annihilation topologies''}
\setcounter{equation}{0}
\label{sec:calc}

\subsection{Definition of the correlator}

The first step in the derivation of the sum rule is the definition of a suitable correlator that allows the extraction of the required matrix element in (\ref{eq:had-ffs}).
To this end, we replace the light baryon in the final state by an interpolating current, 
\begin{align}
    J_\Lambda (y) 
    \equiv 
    \epsilon_{ijk}
    \left[
        u^{T,i}(y) \, C\gamma_5 \slashed{n}_+  d^j(y)      
    \right] 
    \proj
    s^k(y)
    \,,
\end{align}
for which we use the same expression as in Ref.~\cite{Feldmann:2011xf}. 
In particular, we use the projector $\frac{\slashed n_- \slashed n_+}{4}$ to project onto the leading spinor components for a collinear fermion.
Here $C$ is the charge conjugation matrix, the indices $i,\,j,\,k$ are colour indices, and the quark field $u^{T,i}$ denotes the transpose in Dirac space of the field $u^i$.
In the following, the appearance of the charge conjugation matrix is always understood in the chiral representation for Dirac matrices, where 
$$
 C= i\gamma^2\gamma^0 = -C^{-1} = - C^T \qquad \mbox{\small (chiral representation)}\,,
$$
and
\begin{align*}
    &
    C\gamma^\mu C^{-1}= - (\gamma^\mu)^T \,,
    &&
    C\gamma_5 C^{-1}= (\gamma_5)^T \,.
    &
\end{align*}
With this, we define the correlator as 
\begin{align}
    \Pi_\mu(n_- \cdot p') \equiv 
    (i)^2 \int d^4x\, e^{i q\cdot x}
    \int d^4y\, e^{i p'\cdot y}
    \bra{0}
    \T\! \left\{
        J_\Lambda(y) \, 
        \cO_{3-6}(0) \, 
        J_\mu^{\rm em}(x)
    \right\}
    \ket{\Lambda_b(p)}
    \,.
    \label{eq:corr}
\end{align}
Here we consider the correlator as a function of the \emph{small} light-cone projection $n_-\cdot p'$ for a fixed value of the large light-cone projection 
\begin{align}
    n_+\cdot p' \simeq M_{\Lambda_b}-q^2/M_{\Lambda_b}
    \,.
    \label{eq:npp_vs_q2}
\end{align}
As a consequence, the value of $q^2$ is determined by the value of $n_+\cdot p'$ and vice versa.
For the perturbative calculation of the correlator, we take the momentum associated to the interpolating current 
to be \emph{hard-collinear}:
\begin{eqnarray}
 p' \sim (1,0,\lambda^2) \,, \qquad  \mbox{with $p'^2 = m_\Lambda^2 \sim \lambda^2$ \, (hard-collinear)} \,.
\end{eqnarray}
Note that the correlator has an open spinor index inherited from the strange-quark field in the interpolating current $J_\Lambda(y)$.

\subsection{Hadronic representation of the correlator}

To derive the hadronic dispersive representation of the correlator, we use unitarity, which consists in inserting a complete set of hadronic states between the interpolating current and the four-quark operators in \refeq{corr}.
We obtain
\begin{align}
    \!\!\!\!\!\!\!
    \Pi_\mu^\had(n_- \cdot p') 
    & =
    i\int d^4x\, e^{i q\cdot x}
    \sum_{s'} 
    \frac{
        \bra{0}
        J_\Lambda(0)
        \ket{\Lambda(p',s')}
        \bra{\Lambda(p',s')}
        \T\! \left\{
        \cO_{3-6}(0)
        J_\mu^{\rm em}(x)
    \right\}
    \ket{\Lambda_b(p)} 
    }{
        m_\Lambda^2 - (p')^2
    }
    + \dots \!\!\!\!\!\!\!\!\!\!\!\!\!\!\!\!\!\!\!\!\!
    \cr 
    &= 
    \sum_{s'} 
    \frac{
        \bra{0}
        J_\Lambda(0)
        \ket{\Lambda(p',s')}
        \H_\mu^{(i)} 
    }{
        m_\Lambda^2 - (p')^2
    }
   + \dots ,
   \label{eq:Pihad}
\end{align}
where the ellipses denote the contribution of the continuum and excited states and we have used our definition of the non-local matrix elements in \refeq{had-ffs}.
We define the decay constant of the $\Lambda$ baryon as in Ref.~\cite{Feldmann:2011xf}:
\begin{align}
    \langle 0 | 
     J_\Lambda(0) |\Lambda(p',s')\rangle &= 
    (n_+\cdot p') \, f_\Lambda \,
    \proj
    u_\Lambda(p',s') \,,
\label{eq:decay-const}
\end{align}
which implies that $f_\Lambda$ has mass dimension $2$.
Using \refeq{decay-const} and the projections in \refeq{Hperp} yields
\begin{align}
\label{eq:hadrep-p}
    p^\mu\Pi_\mu^\had(n_- \cdot p') 
    & =
    - 
    \frac{M_{\Lambda_b}^2}{2 q^2}
    \frac{f_\Lambda (n_+\cdot p' + m_\Lambda) (n_+\cdot p')}{m_\Lambda^2 - (n_-\cdot p')(n_+\cdot p')} 
    \frac{\slashed{n}_-}{2}
    \nonumber\\
    & \times \!
    \left[
        s_- (M_{\Lambda_b}+ m_\Lambda)\,\H_+^{(i)}(q^2)
        - s_+ (M_{\Lambda_b} - m_{\Lambda})
        \H_{+5}(q^2) \gamma_5
    \right]
    u_{\Lambda_b}(p)
   + \dots ,
    \\
\label{eq:hadrep-g}
    g_\perp^{\mu\nu} \Pi_\mu^\had(n_- \cdot p') 
    &=
    M_{\Lambda_b}^2
    \frac{f_\Lambda (n_+\cdot p' - m_\Lambda)(n_+\cdot p') }{m_\Lambda^2 - (n_-\cdot p')(n_+\cdot p')} 
    \frac{\slashed{n}_-}{2}
    \cr
    & \times
    \left[
    \H_\perp^{(i)}(q^2) 
     \, \gamma^\nu_\perp 
    -  \H_{\perp5}^{(i)}(q^2) 
    \gamma^\nu_\perp  
    \gamma_5 
    \right] u_{\Lambda_b}(p)
   + \dots .
\end{align}
Here in performing the spin summation, we have used that
\begin{equation}
\begin{aligned} 
 \frac{\slashed n_-\slashed n_+}{4} \left( \slashed p'+ m_\Lambda \right)
 u_{\Lambda_b}(p) &= \left( (n_+\cdot p') + m_\Lambda \right) \frac{\slashed n_-}{2} \, u_{\Lambda_b}(p)
 \,, \cr 
  \frac{\slashed n_-\slashed n_+}{4} \left( \slashed p'+ m_\Lambda \right)
\gamma_\perp^\nu \,  u_{\Lambda_b}(p) &= \left( (n_+\cdot p') - m_\Lambda \right) \frac{\slashed n_-}{2} \gamma_\perp^\nu \, u_{\Lambda_b}(p)
\end{aligned}
\end{equation}
in our chosen reference frame with $p_\perp'=0$ and $\frac{\slashed n_- \slashed n_+}{4} \, u_{\Lambda_b}(p) = \frac{\slashed n_-}{2} \, u_{\Lambda_b}(p)$.

\subsection{OPE analysis of the correlator}

\begin{figure}[t!]
	\begin{center} 
	($a$) \quad 
    \parbox[c]{0.4\textwidth}{\includegraphics[width=0.35\textwidth]{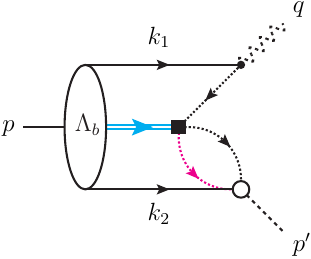}}
	\quad
	($b$)\quad 
    \parbox[c]{0.4\textwidth}{\includegraphics[width=0.35\textwidth]{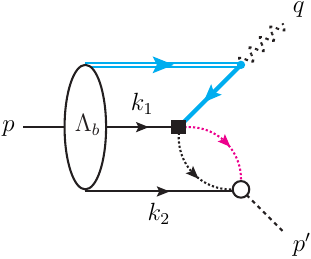}}
    \\[1em]
	($c$) \quad 
    \parbox[c]{0.4\textwidth}{\includegraphics[width=0.35\textwidth]{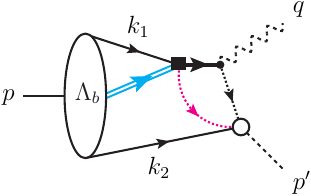}}
    \quad 
	($d$) \quad 
    \parbox[c]{0.4\textwidth}{\includegraphics[width=0.35\textwidth]{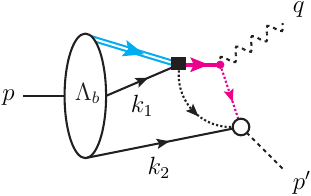}}
    \end{center} 
    \caption{Leading-order annihilation diagrams contributing to the correlator in \refeq{corr}. In each diagram, the lines labelled with momentum $k_1$ and $k_2$ should be identified with up- and down-quark, or vice versa. 
    The $b$-quark is indicated by a light blue line; the strange quark by a magenta line. 
    In addition, (anti-)hard-collinear propagators are indicated by dotted lines, and hard propagators by thick lines.
    The double line denotes the $b$-quark in HQET.
   }
    \label{fig:lcsr-corr}
\end{figure}

For the power counting defined in~\refsec{pow}, the correlator~\eqref{eq:corr} can be calculated using an operator product expansion (OPE).
At leading order in the strong coupling, and restricting ourselves to the ``annihilation topologies'', the OPE analysis 
of the correlator corresponds to the four types of diagrams shown in \reffig{lcsr-corr}. Each diagram comes in two copies, related by isospin symmetry, where the lines labelled by the momenta $k_1$ (and $k_2$) correspond to $u$ (and $d$), or vice versa. Throughout we take the light quarks $u,d,s$ to be massless. 
Notice that the diagram in which the photon is emitted from the spectator quark labelled by $k_2$ does \emph{not} contribute to the discontinuity of $\Pi(n_-\cdot p')$ and can therefore be ignored.

In \reffig{lcsr-corr} we also indicate the virtuality of the internal propagators, which follows from the kinematics given above, together with a method-of-regions analysis of the loop integral. 
As can be seen, diagrams~($b$)-($d$) all contain a \emph{hard} quark propagator, in addition to a loop with two hard-collinear light-quark propagators. In contrast, diagram~($a$) contains an additional anti-hard-collinear propagator instead.
This leads to an enhancement of diagram~($a$) compared to the other three diagrams. After integrating out the hard propagators, diagram~($a$) would thus match onto a correlator in SCET where the quark fields in QCD are trivially replaced by their SCET counterparts. On the other hand, in SCET diagrams~($b$)-($c$) would correspond to another type of correlator in terms of an effective operator involving four quark fields and one additional photon field.
In the following, we thus concentrate on diagram~($a$) since it is expected to give the highest power in the $\Lambda/m_b$ expansion. Moreover, as we discuss in detail below, the appearance of an anti-hard-collinear propagator together with a hard-collinear loop implies that the information about the configuration of the soft momenta $k_1$ and $k_2$ in the $\Lambda_b$ bound state is contained in a new type of soft functions that requires a generalisation of the concept of LCDAs for the $\Lambda_b$. 
This makes diagram~($a$) particularly interesting from a conceptual point of view.

By inserting the effective operators and the currents into \refeq{corr}, we obtain the following expression for diagram~($a$) 
\begin{align}
    \Pi_\mu^\OPE(n_-\cdot p')  
    & =  
    - Q_u \, \epsilon_{\ara{ijk}} \,
    \int d^4x\, e^{i q\cdot x}
    \int d^4y\, e^{i p'\cdot y}
    \nonumber\\*
    & \times
    \bra{0}
    \T \Bigg\{
        \left[
        u^{T,\ara{i}}(y)C\gamma_5 \slashed{n}_+ d^{\ara{j}}(y) 
    \right] 
    \proj
    s^{\ara{k}}(y) 
    \nonumber\\*
    &\times 
        \left[\bar{s}^{\ara{l(m)}} \gamma^\nu P_L b^{\ara{l}}  \right]\!(0)
        \left[\bar{u}^{\ara{m(l)}} \gamma_\nu P_{L(R)} u^{\ara{m}}  \right]\!(0)
        \left[\bar{u}^{\ara{n}} \gamma_\mu u^{\ara{n}} \right]\!(x)
    \Bigg\}
    \ket{\Lambda_b(p)}
    \,,
\end{align}
where colour indices are explicit, and spinor indices within square brackets are contracted.
For concreteness, we have taken the case where the photon is emitted from the $u$ quark. 
Due to the isospin symmetry, the case where the photon is emitted from the $d$ quark gives the same result except for the replacement $Q_u \to Q_d$, since we neglect the light quark masses.
The projector  $P_L$ ($P_R$) acting on the up quark field is for the case of $\cO_3$ and $\cO_4$ ($\cO_5$ and $\cO_6$) operators.
In addition, the colour indices without (with) parenthesis refer to the case of $\cO_3$ and $\cO_5$ ($\cO_4$ and $\cO_6$) operators, respectively.

Performing the Wick contractions corresponding to the diagram \reffig{lcsr-corr}($a$), we get
\begin{equation}
\begin{aligned}
\label{eq:step2}
&  \Pi_\mu^\OPE(n_-\cdot p')  = 
    -  Q_u \, \epsilon_{\ara{ijk}} \, \delta^{\ara{im(il)}} \, \delta^{\ara{kl(km)}} \, \int d^4x\, e^{i q\cdot x}
    \int d^4y\, e^{i p'\cdot y}
    \\
    &\bra{0} \!
         \left[ d^{T,\ara{j}}(y)
        \, C\gamma_5 \slashed{n}_+ \,  S_F^{(u)}(y) \, \gamma_\nu P_{L(R)}\, S_F^{(u)}(-x) \, \gamma_\mu u^{\ara{m}}(x) \right]
   \! \left[
    \proj \,
    S_F^{(s)}(y)
        \, \gamma^\nu P_L b^{\ara{l}}(0) \right] \!
    \ket{\Lambda_b(p)}
    , \!\!\!\!
\end{aligned}
\end{equation}
where the fermion propagators $S_F$ is defined as 
\begin{align}
    S_F (x-y) 
    \equiv
    \bra{0} \T \psi (x) \bar \psi (y) \ket{0}
    = i \int \frac{d^4 p}{(2\pi)^4} \frac{\slashed p + m}{p^2 - m^2 + i\epsilon}
    e^{-ip\cdot(x-y)}
    \,.
\end{align}
Realising that the alternative colour contractions in brackets only yield an extra minus sign and performing the trivial integrations of the space-time coordinates,  \refeq{step2} can be written as
\begin{align}
     \Pi_\mu^\OPE(n_-\cdot p')  
    & = 
    \ara{\pm} i\, Q_u \epsilon_{ijk} 
    \int \frac{d^4\ell_2}{(2\pi)^4} 
    \int \frac{d^4 k_1}{(2\pi)^4} 
    \int \frac{d^4 k_2}{(2\pi)^4}
    \nonumber\\
    &  \times
    \bra{0}
         \left[
            \tilde d^{T,j}(k_2)  C\gamma_5 \slashed{n}_+ 
    \, 
    \frac{\slashed{\ell}_2}{\ell_2^2+i\epsilon}
    \gamma_\nu P_{L(R)}
    \frac{\slashed{k}_1-\slashed q}{(k_1-q)^{2}+i\epsilon}
    \gamma_\mu
    \,
         \tilde u^i(k_1)
    \right] 
    \cr & \times 
     \left[
    \proj  \,
    \frac{\slashed p'-\slashed{\ell}_2-\slashed k_2}{(p'-\ell_2-k_2)^2+i\epsilon}
    \gamma^\nu P_L
   b^k(0)   \right] 
    \ket{\Lambda_b(p)}
    \,,
    \label{eq:step3}
\end{align}
where we denoted the Fourier transformed quark fields with a tilde.

At this step, we perform the tensor reduction for the
integration over the loop momentum $\ell_2$. We define
\begin{eqnarray}
    T^{\alpha\beta} &=& \int [d\ell_2] \, \frac{\ell_2^\alpha \, (q'-\ell_2)^\beta}{(\ell_2^2 +i\epsilon)((q'-\ell_2)^2 +i\epsilon)}
    = A(q'{}^2) \, g^{\alpha\beta} + B(q'{}^2) \, \frac{q'{}^\alpha \, q'{}^\beta}{q'{}^2} 
\end{eqnarray}
with $q'=p'-k_2$ and --- for the moment --- consider the loop integration in $D\neq 4$ dimensions, such that
$$
[d\ell_2] \equiv i \, \mu^{4-D} \, \frac{d^D\ell_2}{(2\pi)^D} \,.
$$
Since the LCSRs are derived using a dispersion relation of the form
\begin{align}
    \label{eq:disrel}
    \Pi_\mu^\OPE(n_-\cdot p') =
    \frac{1}{\pi} \, \int\limits_0^\infty ds \, \frac{\Im  \, \Pi_\mu^\OPE(s/(n_+\cdot p'))}{s- (n_+\cdot p')(n_-\cdot p') - i\epsilon} \,, \qquad n_+\cdot p' > 0 \,,
\end{align}
we only need the discontinuity of the correlator.
Thus, only the imaginary part of the integrals $A(s)$ and $B(s)$ is relevant in our calculation, which is finite for $D \to 4$.
Performing the algebra, we have 
\begin{align}
    A(q'{}^2) &= \frac{1}{D-1} \, \frac{1}{q'{}^2} \,  \int [d\ell_2] \, \frac{(q'\cdot \ell_2)^2 - \ell_2^2 \, q'{}^2}{(\ell_2^2 +i\epsilon)((q'-\ell_2)^2 +i\epsilon)} \,,
    \\[0.2cm]
    B(q'{}^2) &= \frac{1}{q'{}^2} \,  \int [d\ell_2] \, \frac{\left( q'{}^2- (\ell_2 \cdot q')\right) (\ell_2\cdot q')}{(\ell_2^2 +i\epsilon)((q'-\ell_2)^2 +i\epsilon)} - A(q'{}^2)
    \,.
\end{align}
The imaginary part of these integrals is then easily calculated:
\begin{align}
    &
    \Im A(s) = 
    -\frac{s \, \theta[s]}{192\pi} \,,
    &&
    \Im B(s) = 
    -\frac{s \, \theta[s]}{96\pi} \,.
    &
\end{align} 
Inserting these results in \refeq{step3}, we obtain
\begin{align}
  \Im  \, \Pi_\mu^\OPE(n_-\cdot p')  
    & =
    \ara{\mp} \frac{Q_u \epsilon_{ijk}}{192\pi} 
    \int \frac{d^4 k_1}{(2\pi)^4} 
    \int \frac{d^4 k_2}{(2\pi)^4} \, \theta((p'-k_2)^2)
    \nonumber 
    \\ &  \times
    \left\{ g^{\alpha\beta} \, (p'-k_2)^2 + 
    2 \, (p'-k_2)^\alpha (p'-k_2)^\beta\right\}
    \nonumber\\
    &  \times
    \bra{0}
         \left[
            \tilde d^{T,j}(k_2)  C\gamma_5 \slashed{n}_+ 
    \, 
    \gamma_\alpha 
    \gamma_\nu P_{L(R)}
    \frac{\slashed{k}_1-\slashed q}{(k_1-q)^{2}+i\epsilon}
    \gamma_\mu
    \,
         \tilde u^i(k_1)
    \right] 
    \cr &  \times 
     \left[
    \proj 
    \gamma_\beta 
    \gamma^\nu P_L
   b^k(0)   \right] 
    \ket{\Lambda_b(p)} \,.
    \label{eq:corr_tmp1}
\end{align} 
We observe that at leading power in the expansion parameter $\lambda$, the above expression depends  on two \emph{opposite} light-cone projections of the light-quark momenta in the $\Lambda_b$ baryon. 
The integrals over the loop momentum $\ell_2$ depend on 
$$ (p'-k_2)^2 \simeq (n_+ \cdot p')( n_-\cdot p'-n_- \cdot k_2) \sim \lambda^2 \,, $$ while the remaining light-quark propagator connecting the external photon and the effective four-quark operator depends on 
$$ (k_1-q)^2 \simeq (n_- \cdot q)(n_+\cdot q- n_+ \cdot k_1) \sim \lambda^2 \,. $$ 
In the SCET jargon, the associated short-distance dynamics is described by \emph{hard-collinear} modes, but in different light-cone directions (one in the direction of $p'$, and the other in the direction of $q$).
After performing a Fourier transform to position space, this would correspond to tri-local matrix elements of the form 
$$
  \langle 0 | d^{j}_\beta(\tau_2 n_-) \, u^{i}_\alpha(\tau_1 n_+) \, b^{k}_\delta(0) |\Lambda_b(p)\rangle \,, 
$$
where we have made the spinor indices $\alpha,\beta,\delta$ explicit.
These objects define a new type of three-particle LCDAs or, more appropriately, ``soft functions'' in the context of SCET factorisation (for this, the light-quark fields have to be supplemented with the corresponding soft Wilson lines to render the tri-local operator gauge invariant). 
Their appearance in the ``annihilation topologies'' is due to the fact that the two quarks take part in different dynamics: one, associated to the interaction with an anti-collinear photon; and another, associated with the collinear momentum of the interpolating current.\footnote{The appearance of soft $b$-hadron matrix elements with two light-like separations has already been observed in other contexts, see e.g. Refs.~\cite{Chay:2007ej,Benzke:2010js,Kozachuk:2018yxf,Beneke:2022msp,Qin:2022rlk,Piscopo:2023opf}.}
Before proceeding with the sum rule, we discuss and classify the new type of soft functions and the connection with the conventional baryon LCDAs.

\subsection[$\Lambda_b$ soft functions with two light-like separations]{$\boldsymbol{\Lambda_b}$ soft functions with two light-like separations}

The standard definitions of three-particle LCDAs for the  $\Lambda_b$ baryon can be found in Ref.~\cite{Ball:2008fw}.
In order to generalise these definitions to our case,
it is convenient to follow the procedure of Ref.~\cite{Bell:2013tfa}, where  the starting point is the decomposition of the hadronic matrix element of a general tri-local operator:
\begin{multline}
  \epsilon_{ijk} \, \langle 0| \left(u^i_\alpha (z_1) \,  d^j_\beta(z_2)\right) b^k_\delta(0)
  |\Lambda_b(p)\rangle \equiv
  \\* 
  \frac14 \left \{ f^{(1)}_{\Lambda_b} \left[ \tilde M^{(1)}(v,z_1,z_2) \gamma_5 C^{-1}\right]_{\beta\alpha}
  +
f^{(2)}_{\Lambda_b} \left[ \tilde M^{(2)}(v,z_1,z_2) \gamma_5 C^{-1}\right]_{\beta\alpha} \right\} u_{\Lambda_b,\delta}(v) \,.
\label{eq:LCDAgeneral}
\end{multline}
Here $z_1$ and $z_2$ are space-time points and QCD gauge-links are understood implicitly.
The Dirac matrices $\tilde M^{(1(2))}$ contain an even (odd) number of Dirac matrices, respectively.
In the following, we focus on the matrix $\tilde M^{(2)}$ which is relevant for our sum rule.
(The matrix $\tilde M^{(1)}$ can be treated in a completely analogous way, see Ref.~\cite{Bell:2013tfa}.) 
Here, the most general Lorentz-covariant decomposition is
\begin{align}
\tilde M^{(2)}(v,z_1,z_2) &=
\slashed v \, \tilde \Phi_2(t_1,t_2,z_1^2,z_2^2,z_1 \cdot z_2)
+
\frac{\tilde \Phi_X(t_1,t_2,z_1^2,z_2^2,z_1\cdot z_2)}{4 t_1 t_2} \left( \slashed z_2 \slashed v \slashed z_1 - \slashed z_1 \slashed v \slashed z_2 \right)
\nonumber\\*  
&+
\frac{\tilde \Phi_{42}^{(i)}(t_1,t_2,z_1^2,z_2^2, z_1\cdot z_2)}{2t_1} \, \slashed z_1
+
\frac{\tilde \Phi_{42}^{(ii)}(t_1,t_2,z_1^2,z_2^2,z_1 \cdot z_2)}{2t_2} \, \slashed z_2 \,,
\label{eq:barygeneral1}
\end{align}
where $t_i =v\cdot z_i$. 
The standard LCDAs can be obtained by expanding the above expression around the limit $z_1^2=z_2^2=z_1\cdot z_2=0$.
However, for the new type of soft functions, we rather have to consider the expansion for 
the situation 
$n_+\cdot z_1 \ll z_1^\perp \ll n_-\cdot z_1$ and $n_-\cdot z_2 \ll z_2^\perp \ll n_+\cdot z_2$, 
such that $t_1 \approx \bar \tau_1 = \frac{n_- \cdot z_1}{2}$, $t_1 \approx \tau_2 = \frac{n_+ \cdot z_2}{2}$, and 
$z_1 \cdot z_2 \approx 2 \bar \tau_1 \tau_2$.
This yields
\begin{equation}
\begin{aligned}
\tilde M^{(2)}(v,z_1,z_2) & \longrightarrow
\frac{\slashed n_+}{2} 
\left(
\tilde \chi_2(\bar{\tau}_1,\tau_2)
+
\tilde \chi_{42}^{(i)}(\bar{\tau}_1,\tau_2)
\right)
+
\frac{\slashed n_-}{2} \left( \tilde \chi_2(\bar{\tau}_1,\tau_2)+\tilde \chi_{42}^{(ii)}(\bar{\tau}_1,\tau_2) \right)
\\
& \quad
+
\frac{\tilde \chi_{42}^{(i)}(\bar{\tau}_1,\tau_2)}{2\bar{\tau}_1} \,  \slashed z_1^\perp
+
\frac{\tilde \chi_{42}^{(ii)}(\bar{\tau}_1,\tau_2)}{2\tau_2} \,  \slashed z_2^\perp
\\
& \quad
+
\tilde\chi_X(\bar{\tau}_1,\tau_2) \left( \frac{\slashed z_1^\perp}{2\bar{\tau}_1} + \frac{\slashed z_2^\perp}{2\tau_2}
 \right) 
 \!
 \left(\frac{\slashed n_- \slashed n_+}{4} - \frac{\slashed n_+ \slashed n_-}{4} \right)
 + \cO(z_{i\perp}^2, n_-\!\cdot z_2 ,  n_+\!\cdot z_1) \,,
 \!\!
 \label{eq:M2zexp}
\end{aligned}
\end{equation}
where $\bar\tau_1 =n_-\cdot z_1/2$ and $\tau_2 =n_+\cdot z_2/2$, and 
\begin{align*}
    \tilde \chi_2(\bar \tau_1,\tau_2) &= \tilde \Phi_2(\bar \tau_1,\tau_2,0,0,2 \bar\tau_1 \tau_2) \qquad \mbox{etc.}
\end{align*}
This should be compared with the standard LCDAs, which are given by 
$$  \tilde \phi_2(\tau_1,\tau_2) = \tilde \Phi_2(\tau_1,\tau_2,0,0,0) 
\qquad \mbox{etc.} 
$$
We introduce the LCDAs in momentum space by performing a Fourier transform
\begin{align}
    \chi (\bar{\omega}_1,\omega_2) \equiv 
    \int\limits_{-\infty}^\infty
    \frac{d \bar{\tau}_1}{2\pi} \, e^{i\bar{\omega}_1\bar{\tau}_1} 
    \int\limits_{-\infty}^\infty
    \frac{d \tau_2}{2\pi} \, e^{i \omega_2 \tau_2} \,
    \tilde \chi (\bar \tau_1,\tau_2)
\end{align}
Therefore, from the Fourier transform of \refeq{M2zexp}, 
we can construct the momentum-space projector:
\begin{align}
    M^{(2)}(\bar{\omega}_1,\omega_2) 
    & =   
    \frac{\slashed n_+}{2} \, 
    \left(
        \chi_2(\bar{\omega}_1,\omega_2)
        +
        \chi_{42}^{(i)}(\bar{\omega}_1,\omega_2)
    \right)
    \cr
    & +
    \frac{\slashed n_-}{2} 
    \left( 
        \chi_2(\bar{\omega}_1,\omega_2)
        +
        \chi_{42}^{(ii)}(\bar{\omega}_1,\omega_2) 
    \right)
    \cr
    & - 
    \frac{1}{2} \, \gamma_\rho^\perp
    \left( 
        \bar\chi_{42}^{(i)}(\bar{\omega}_1,\omega_2) - \bar\chi_X(\bar{\omega}_1,\omega_2) 
    \right)
    \frac{\slashed n_+\slashed n_-}{4}
    \, \frac{\partial}{\partial k_{1\rho}^\perp}
    \cr 
    & -
    \frac{1}{2} \, \gamma_\rho^\perp
    \left( 
        \bar\chi_{42}^{(i)}(\bar{\omega}_1,\omega_2) + \bar\chi_X(\bar{\omega}_1,\omega_2) 
    \right)
    \frac{\slashed n_-\slashed n_+}{4}
    \, \frac{\partial}{\partial k_{1\rho}^\perp}
    \cr 
    & - 
    \frac{1}{2} \, \gamma_\rho^\perp 
    \left( 
        \hat\chi_{42}^{(ii)}(\bar{\omega}_1,\omega_2) + \hat\chi_X(\bar{\omega}_1,\omega_2) 
    \right)
    \frac{\slashed n_-\slashed n_+}{4}
    \, \frac{\partial}{\partial k_{2\rho}^\perp}
    \cr
    & -
    \frac{1}{2} \, \gamma_\rho^\perp
    \left( 
        \hat\chi_{42}^{(ii)}(\bar{\omega}_1,\omega_2) - \hat\chi_X(\bar{\omega}_1,\omega_2) 
    \right)
    \frac{\slashed n_+\slashed n_-}{4}
    \, \frac{\partial}{\partial k_{2\rho}^\perp} \,,
    \label{eq:newlcproj}
\end{align}
where the derivatives are understood
to act on a hard-scattering kernel that has been Taylor-expanded in transverse momenta, and we define
\begin{align}
    & \bar \omega_1 = n_+ \cdot k_1 \,, \qquad 
    \omega_2 = n_- \cdot k_2 \,. &
\end{align}
We also introduced the abbreviations 
\begin{equation}
\begin{aligned}
    \bar{\chi}(\bar{\omega}_1,\omega_2) & \equiv \int_0^{\bar{\omega}_1} d\bar\eta_1 \, \chi(\bar\eta_1,\omega_2)
    \,,\\
    \hat{\chi}(\bar{\omega}_1,\omega_2) & \equiv \int_0^{\omega_2} d\eta_2 \, \chi (\bar{\omega}_1,\eta_2)
    \,.
\end{aligned}
\end{equation}
Explicit models for these LCDAs are derived in \refapp{LCDAmodel}.

\subsection{Expressing the correlator in terms of soft functions}
\label{sec:finalOPE}

We can now proceed to express the hadronic matrix element appearing in the leading order result for  the correlator (\ref{eq:corr_tmp1}) in terms of the momentum-space projector following from (\ref{eq:LCDAgeneral}) and (\ref{eq:newlcproj}).
To this end, we write the hadronic matrix element in \refeq{corr_tmp1} as a Dirac trace: 
\begin{align}
\label{eq:trace}
& 
\int \frac{d^4 k_1}{(2\pi)^4} 
    \int \frac{d^4 k_2}{(2\pi)^4} \,
\epsilon_{ijk} \, \langle 0 | \left[ \tilde d^{T,j}(k_2) \, \Gamma \, \tilde u^i(k_1) \right] b^k(0) |\Lambda_b(p)\rangle \, f^{\alpha\beta}(k_2^\perp)
\cr 
=&
-
\int \frac{d^4 k_1}{(2\pi)^4} 
    \int \frac{d^4 k_2}{(2\pi)^4} \,
\epsilon_{ijk} \, \langle 0 | \left[ \tilde u^{T,i}(k_1) \, \Gamma^T \, \tilde d^j(k_2) \right] b^k(0) |\Lambda_b(p)\rangle \, f^{\alpha\beta}(k_2^\perp)
\cr 
=&
 -  \frac14 \, f_{\Lambda_b}^{(2)} \,
\int\limits_0^\infty d\bar\omega_1 \, \int\limits_0^\infty d\omega_2  
  \, 
 {\rm tr}[M^{(2)} (\bar{\omega}_1,\omega_2) \gamma_5 C^{-1} \, \Gamma^T] \,
 f^{\alpha\beta}(k_2^\perp)\, u_{\Lambda_b}(v) \Big|_{k_i^\perp=0} \,, 
\end{align}
with
\begin{align} 
    \Gamma=
   C\gamma_5 \, \slashed{n}_+ 
    \, 
    \gamma_\alpha 
    \gamma_\nu P_{L(R)}
    \left(\slashed{k}_1-\slashed q\right)
    \gamma_\mu 
    \,
\end{align}
and hence
\begin{align}
\label{eq:GammaT}
    \Gamma^T = C \gamma_\mu (\slashed k_1-\slashed q) \, P_{L(R)} \,\gamma_\nu \gamma_\alpha \slashed n_+ \gamma_5 \,.
\end{align}
The function $f^{\alpha\beta}(k_2^\perp)$ can be read off \refeq{corr_tmp1}:
\begin{align}
    \label{eq:fexp}
    f^{\alpha\beta}(k_2^\perp) & \equiv
    g^{\alpha\beta} \, (p'-k_2)^2 + 2 \, (p'-k_2)^\alpha (p'-k_2)^\beta 
    \cr 
    & = \frac{(n_+ \cdot p')^2}{2}\, n_-^\alpha n_-^\beta  + (n_+\cdot p')(n_-\cdot p'-n_- \cdot k_2) \, g_\perp^{\alpha\beta} 
    \cr  & \quad {}
     - (n_+\cdot p') \, n_-^\alpha (k^\perp_2)^\beta 
     - (n_+\cdot p') \, n_-^\beta (k^\perp_2)^\alpha +\order{\lambda^4} \,,
\end{align}
where we have kept terms of order $\lambda^2$ but dropped subsubleading terms of order $\lambda^4$,
as well as terms proportional to $n_+^\alpha$ or $n_+^\beta$ which do  not contribute in (\ref{eq:corr_tmp1}).
To derive \refeq{GammaT} we have used that $\gamma_\mu^T = - C  \gamma^\mu  C^{-1}$, and the additional sign in front of the trace stems from anti-commuting the field operators $u_\alpha(z_1)$ and $d_\beta(z_2)$ when using eq.~(\ref{eq:LCDAgeneral}) in (\ref{eq:corr_tmp1}). 
As already mentioned, the projector $M^{(1)}$ does not contribute to the trace, because the Dirac structure $\Gamma$ has an odd number of Dirac matrices.
For later convenience, we also expand the term $(\slashed k_1-\slashed q)$ up to $\order{\lambda^4}$:
\begin{align}
    \slashed k_1-\slashed q 
    & = 
    - (n_- \cdot q) \, \frac{\slashed n_+}{2}
    \nonumber\\
    & \qquad {} +
     \slashed{k}_1^{\perp }+ \frac{\slashed{n}_+}{2}  (n_- \cdot k_1)  + \frac{\slashed{n}_-}{2}  (\bar{\omega}_1 - n_+ \cdot q) +\order{\lambda^4}
     \,.
\end{align}

Let us first consider the contribution arising from the leading term in $f^{\alpha\beta}$ together with the leading term in ($\slashed k_1-\slashed q)$. For this we obtain 
\begin{multline}
\label{eq:trLP}
 -  \frac14 \, {\rm tr}[M^{(2)} \gamma_5 C^{-1} \, \Gamma^T] \, f^{\alpha\beta}(k_2^\perp)\Big|_{k_i^\perp=0, \ \rm leading} =
 \\
  (n_-\cdot q) \, (n_+ \cdot p')^2 \, n_-^\beta 
 \left( - \frac{g_{\mu\nu}^\perp}{2} 
 \pm 
 \frac{
    i \epsilon_{\mu\nu\{n_-\}\{n_+\}} }{4}
\right)
  \left( \chi_2(\bar \omega_1,\omega_2) + \chi_{42}^{(ii)}(\bar\omega_1,\omega_2) \right) 
    \,,
\end{multline} 
where the upper sign is for the operators $\cO_{3,4}$, and the lower sign for $\cO_{5,6}$.
For the Levi-Civita symbol we use the convention $\epsilon^{0123}=1$ and the short-hand notation $\epsilon_{\mu\nu\rho\sigma} q^\sigma \equiv \epsilon_{\mu\nu\rho\{q\}}$.
Note that \refeq{trLP} fixes the Lorentz index $\mu$ to be transversal and hence this term only contributes to the LCSR for $\H_{\perp (5)}^{(i)}$.
We can now easily obtain the leading contribution to the OPE calculation of $\Im  \, \Pi_\mu^\OPE$ by plugging \refeqa{trace}{trLP} into \refeq{corr_tmp1}:
\begin{align}
\label{eq:LPImperp}
    \Im  \, \Pi_\mu^\OPE(n_-\cdot p')  =
    & 
    \ara{\mp} \frac{Q_u  f_{\Lambda_b}^{(2)}}{192\pi} \, (n_+ \cdot p')^2 \,
    \nonumber \\* & \times 
    \int\limits_0^\infty d\bar\omega_1 \, \int\limits_0^\infty d\omega_2 \, 
    \frac{
        \theta(n_-\cdot p' - \omega_2)
    }{
        n_+ \cdot q - \bar \omega_1 + i\epsilon
    } 
  \left( \chi_2(\bar \omega_1,\omega_2) + \chi_{42}^{(ii)} (\bar\omega_1,\omega_2) \right) 
    \nonumber \\*
    & \times 
    \left( - \frac{g_{\mu\nu}^\perp}{2} 
    \pm 
    \frac{
    i\epsilon_{\mu\nu\{n_-\}\{n_+\}} }{4}
    \right)
    \left[
    \slashed n_- 
    \gamma^\nu_\perp P_L
    \right] u_{\Lambda_b}(v)
    + \order{\lambda^2}
    \,.
\end{align}
This can be further simplified by using that after combining with the Dirac projection of the $b$-quark field in (\ref{eq:corr_tmp1}), we have 
\begin{align}
    \left( 
        - \frac{g_{\mu\nu}^\perp}{2} +
        \frac{i \epsilon_{\mu\nu\{n_-\}\{n_+\} }
        }{4} 
    \right)
    \left[
        \slashed n_- \gamma^\nu_\perp P_L
    \right]
    & =
    0  \,,
    \label{eq:gid1}
    \\
    \left( 
        - \frac{g_{\mu\nu}^\perp}{2} -
        \frac{i \epsilon_{\mu\nu\{n_-\}\{n_+\}}
        }{4} 
    \right)
    \left[
        \slashed n_- \gamma^\nu_\perp P_L
    \right]
    & =
    - \slashed n_- \gamma_\mu^\perp P_L
    \,,
    \label{eq:gid2}
\end{align}
which holds in $D=4$ dimensions.
Therefore only the operators $\cO_5$ and $\cO_6$ contribute to the LCSR for $\H_{\perp (5)}^{(i)}$ in the considered order of the calculation.

Actually, this result can already be understood by considering the Dirac structure of the original four-quark operators in QCD and projecting the light-quark fields onto the leading collinear or anti-collinear spinor components in SCET that correspond to the topology in Fig.~\ref{fig:lcsr-corr}($a$).
Ignoring the colour indices for the moment, one has by virtue of \refeqa{gid1}{gid2}
\begin{align}
    \label{eq:projO3O4}
    \left(\bar{s} \, \projb \, \gamma^\nu P_L b  \right)
    \left(\bar{q} \, \projb \, \gamma_\nu P_L\, \projb \, q \right)
    & = 0\,,
    \\
    \label{eq:projO5O6}
    \left(\bar{s} \, \projb \,\gamma^\nu P_L b  \right)
    \left(\bar{q} \, \projb \, \gamma_\nu P_R\,  \projb \, q \right)
    & = 
    \left(\bar{s} \, \frac{\slashed n_+ \slashed n_-}{4} \, \gamma^\nu_\perp P_L b  \right)
    \left(\bar{q} \, \gamma^\perp_\nu P_R\, q  \right) 
    \,.
\end{align}
It follows from this argument, that neither the current-current operators $\cO_1^{(u)}$ and $\cO_2^{(u)}$ --- which have larger Wilson coefficients, but enter with Cabibbo-suppressed CKM factors in $b \to s$ transitions --- contribute to the baryonic annihilation topologies at leading power.
It is worth noting that in the mesonic counterpart, the role of $\bar q$ and $q$ is interchanged in the leading annihilation topology compared to the baryonic case; and for that reason -- by the same argument -- only the operators ${\cal O}_3$ and ${\cal O}_4$ (and also $\cO_1^{(u)}$ and $\cO_2^{(u)}$) contribute at leading power \cite{Beneke:2002ph}.

To proceed in the calculation of the associated non-local form factors $\H_+^{(i)}$, we contract \refeq{trace} with $p^\mu$.
Since the leading term vanishes, we consider sub-leading terms of $\order{\lambda^2}$:
\begin{multline}
    \label{eq:trNLP}
    -  \frac14 \,p^\mu \,{\rm tr}[M^{(2)} \gamma_5 C^{-1} \, \Gamma^T] f^{\alpha\beta}(k_2^\perp)\Big|_{k_i^\perp=0, \ \rm subleading} 
    \\
    =
    -\frac{1}{4}  M_{\Lambda_b} (n_+\cdot p') (n_-\cdot q) \!
    \left( 
        \hat\chi_{42}^{(ii)}(\bar{\omega}_1,\omega_2) + \hat\chi_X(\bar{\omega}_1,\omega_2) 
    \right)\!\!
    \left(
        3 n_-^\beta n_+^\nu - 2 g^{\beta\nu}
        \mp i \epsilon^{\beta\nu\{ n_+\}\{n_-\}}
    \right)\!
    ,
\end{multline}
where the terms proportional to $n_-^\beta n_-^\nu$ and $n_+^\beta$ have been dropped, since they vanish once contracted with the Dirac matrix in the last line of \refeq{corr_tmp1}.
We obtain the leading contribution to the OPE calculation of $p^\mu \Im  \, \Pi_\mu^\OPE$ by inserting \refeqa{trace}{trNLP} into \refeq{corr_tmp1} contracted with $p^\mu$:
\begin{align}
\label{eq:LPIpmu}
    p^\mu \Im  \, \Pi_\mu^\OPE(n_-\cdot p')  =
    & 
    \ara{\pm} \frac{1}{4} \frac{Q_u  f_{\Lambda_b}^{(2)} }{192\pi} \, M_{\Lambda_b} (n_+ \cdot p') \,
    \nonumber \\* & \times 
    \int\limits_0^\infty d\bar\omega_1 \, \int\limits_0^\infty d\omega_2 \, 
    \frac{\theta(n_-\cdot p' - \omega_2)}{n_+ \cdot q - \bar \omega_1 + i\epsilon} 
    \left( 
        \hat\chi_{42}^{(ii)}(\bar{\omega}_1,\omega_2) + \hat\chi_X(\bar{\omega}_1,\omega_2) 
    \right)
    \nonumber \\*
    & \times 
    \left[\gamma_\beta \slashed n_- \right]\left( \frac{g^{\beta\nu}_\perp}{2} \pm \frac{i \epsilon^{\beta\nu\{ n_+\}\{n_-\}}}{4} \right)  \left[
        \slashed n_+
        \gamma_\nu P_L
    \right]  u_{\Lambda_b}(v)
    + \order{\lambda^4}
    \,,
\end{align}
where we have used that
\begin{align}
    \label{eq:gid3}
    \left[\gamma_\beta \slashed n_- \right]
    \!\left( \frac{g^{\beta\nu}_\perp}{2} - \frac{i \epsilon^{\beta\nu\{ n_+\}\{n_-\}}}{4} \right) \! \left[
        \slashed n_+
        \gamma_\nu P_L
    \right] 
    & = 0 
    \\
    \label{eq:gid4}
    \left[\gamma_\beta \slashed n_- \right]
    \!\left( \frac{g^{\beta\nu}_\perp}{2} + \frac{i \epsilon^{\beta\nu\{ n_+\}\{n_-\}}}{4} \right)  \!\left[
        \slashed n_+
        \gamma_\nu P_L
    \right] u_{\Lambda_b}(v)
    & = 8 \left[
        \frac{\slashed n_-
        \slashed n_+}{4}
        P_L
    \right] u_{\Lambda_b}(v)=
    8 \left[
        \frac{\slashed n_-
        }{2}
        P_R
    \right] u_{\Lambda_b}(v) \,.
\end{align}
Therefore, in analogy with $\H_{\perp (5)}^{(i)}$ and as a consequence of \refeqa{projO3O4}{projO5O6}, only the operators $\cO_5$ and $\cO_6$ contribute to the LCSR for $\H_{+ (5)}^{(i)}$.

\subsection{Derivation of light-cone sum rules}
\label{sec:LCSR}

Following the usual procedure to derive a LCSR --- see e.g Ref.~\cite{Colangelo:2000dp} --- we match the OPE calculation of the correlator $\Pi_\mu$ of \refeqa{LPImperp}{LPIpmu} onto the corresponding hadronic representations of \refeqa{hadrep-p}{hadrep-g}.
The contribution of the continuum and excited states is removed by using the semi-global quark-hadron duality approximation. 
In practice, we assume that the second line of \refeqs{hadrep-p}{hadrep-g} is equal to the dispersive integral in the OPE calculation above the effective threshold $s_0$, whose value is discussed in \refsec{num}.
We perform a Borel transform with respect to the variable $(n_- \cdot p')$ to further suppress the continuum and excited states contribution.
This reduces the systematic error due to the quark-hadron duality approximation.
Performing a Borel transform in our case consists in replacing
\begin{align}
    \frac{1}{{\cal K} - n_-\cdot p'}
    \to
    e^{-\frac{{\cal K} }{\omega_M}}
    \,,
\end{align}
where $\omega_M$ is the associated Borel parameter and ${\cal K} $ does not depend on $(n_-\cdot p')$.
The bulky formulae resulting from the OPE calculation with the hadronic representation are collected in \refapp{match}.
In our calculation, this matching implies that (taking into account Eqs.~\eqref{eq:gid1}, \eqref{eq:gid2}), \eqref{eq:gid3}, and \eqref{eq:gid4})
\begin{equation}
\begin{aligned}
    \label{eq:relH}
    &
    \H_{\perp(5)}^{(5)} = - \H_{\perp(5)}^{(6)} \,, 
    &&
    \H_{\perp(5)}^{(3)} = \H_{\perp(5)}^{(4)} = 0  \,,
    &&
    \H_{\perp}^{(i)} = \H_{\perp5}^{(i)} \,,
    &
    \\
    &
    \H_{+(5)}^{(5)} = - \H_{+(5)}^{(6)} \,, 
    &&
    \H_{+(5)}^{(3)} = \H_{+(5)}^{(4)} = 0  \,,
    &&
    \H_{+}^{(i)} = - \frac{s_+ (M_{\Lambda_b} - m_{\Lambda})}{ s_- (M_{\Lambda_b}+ m_\Lambda)} \H_{+5}^{(i)} \,,
    &
\end{aligned}
\end{equation}
up to higher order corrections.
Hence, in the limit $M_{\Lambda_b} \gg m_{\Lambda}$, the last of these identities becomes
$
    \H_{+}^{(i)} = - \H_{+5}^{(i)} 
$.
Thus, it is sufficient to present the LCSRs for, e.g., only $\H_\perp^{(5)}$ and $\H_+^{(5)}$.
These LCSRs read
\begin{multline}
    \label{eq:LCSRH5perp}
    \H_\perp^{(5)}(q^2) 
    =
    \frac{f_{\Lambda_b}^{(2)}}{f_\Lambda} 
    \frac{(Q_u + Q_d) }{192\pi^2}
    \frac{1}{ M_{\Lambda_b}^2 (n_+\cdot p' - m_\Lambda)}
    \\*
    \times
    \int\limits_0^{\sigma_0} d\sigma \int\limits_0^\sigma d\omega_2
    \int\limits_0^\infty d\bar\omega_1 \, 
    \frac{(n_+ \cdot p')^2}{n_+ \cdot q - \bar \omega_1 + i\epsilon} 
  \left( \chi_2(\bar \omega_1,\omega_2) + \chi_{42}^{(ii)}(\bar\omega_1,\omega_2) \right) 
    e^{\frac{m_\Lambda^2/(n_+\cdot p') - \sigma}{\omega_M} }
    \,,
\end{multline} 
and 
\begin{multline} 
    \H_+^{(5)}(q^2)
     =
     -
    \frac{f_{\Lambda_b}^{(2)}}{f_\Lambda}  
     \frac{(Q_u + Q_d)  }{192\pi} 
    \frac{q^2}{M_{\Lambda_b} (n_+\cdot p' + m_\Lambda)} 
    \frac{1}{s_- (M_{\Lambda_b}+ m_\Lambda)}
    \\
    \times 
    \int\limits_0^{\sigma_0} d\sigma \int\limits_0^\sigma d\omega_2
    \int\limits_0^\infty d\bar\omega_1 \,
    \frac{n_+\cdot p'}{n_+ \cdot q - \bar \omega_1 + i\epsilon} 
    \left( 
        \hat\chi_{42}^{(ii)}(\bar{\omega}_1,\omega_2) + \hat\chi_X(\bar{\omega}_1,\omega_2) 
    \right)
    e^{\frac{m_\Lambda^2/(n_+\cdot p') - \sigma}{\omega_M} }
    \,,
    \label{eq:LCSRH5plus}
\end{multline}
where $\sigma_{(0)} \equiv s_{(0)}/(n_+ \cdot p')$.
The integrals can be performed analytically, assuming the models for the LCDAs given in \refapp{LCDAmodel}.
These formulae are given in \refapp{match}.

The LCSRs for the non-local form factors $\H_{\perp(5)}^{(i)}$ and $\H_{+(5)}^{(i)}$ can be compared with the LCSR for the local form factor $\xi_\Lambda$ derived in Ref.~\cite{Feldmann:2011xf}.
In the large recoil limit, i.e. $(n_+\cdot p')\sim M_{\Lambda_b}$, $\xi_\Lambda$ is equal to each of the helicity form factors:
\begin{align}
    \label{eq:xirel}
    \xi_\Lambda (q^2)
    \simeq f_0 (q^2) 
    \simeq f_+ (q^2) 
    \simeq f_\perp (q^2) 
    \simeq g_0 (q^2) 
    \simeq g_+ (q^2) 
    \simeq g_\perp (q^2) 
    \,.
\end{align}
The LCSR for $\xi_\Lambda$ reads
\begin{align}
    \label{eq:xiexpl}
    \xi_\Lambda (q^2)
    =
    \frac{f_{\Lambda_b}^{(2)}}{f_\Lambda} 
    \frac{1}{(n_+\cdot p')}
    \int\limits_0^{\sigma_0} d\sigma \,
    \phi_4(\sigma) \,
    e^{\frac{m_\Lambda^2/(n_+\cdot p') - \sigma}{\omega_M} }\,,
\end{align}
where $\phi_4$ is one the standard LCDAs partially integrated (see Ref.~\cite{Feldmann:2011xf} for its definition).
Comparing this sum rule with the one for $\H_\perp^{(5)}$, we observe that they contribute at the same power of $\lambda^2$.
The factor of $192\pi$ due to the loop in \refeq{LCSRH5perp} is compensated in the decay amplitude by the factor $- 16\pi^2 \,  \frac{2M_{\Lambda_b}^2}{q^2}$ appearing in \refeqs{deltaC9a}{deltaC9d}.
Therefore the suppression of the ``annihilation topologies'' is only due the small Wilson coefficients of the operators $\cO_{3-6}$.
It is also important to stress that, while local form factors are real-valued, the non-local form factors are generally complex-valued.
This is evident in our LCSRs, as there is a pole in the integration path of $d\bar{\omega}_1$.
From a phenomenological point of view, this imaginary part is due to $u\bar u$ and $d\bar d$ hadronic states going on shell.
In other words, $\H_{\perp(5)}^{(i)}$ and $\H_{+(5)}^{(i)}$ have a branch cut on the real positive axis starting at $q^2 =4m_\pi^2$.

It is also interesting to compare our results for $\Lambda_b \to \Lambda \ell^+\ell^-$ decays calculated using LCSRs with the  corresponding results for $B\to K^{(*)} \ell^+\ell^-$ calculated using QCD factorisation, i.e. the annihilation topologies~\cite{Bobeth:2007dw,Beneke:2001at}.
Confronting our \refeqs{LCSRH5perp}{LCSRH5plus} with Eq.~(18) of Ref.~\cite{Beneke:2001at}, we find that the two hard scattering kernels have a very similar structure, with a pole appearing in the denominator for any $q^2>0$.
Another analogy concerns the fact that in both the mesonic and baryonic cases the leading contribution comes from the diagram where the photon is emitted from the spectator quark in the $b$ hadron.
However, as mentioned above, in the mesonic case the contributing operators are $\cO_3$ and $\cO_4$, while in the baryonic case they are $\cO_5$ and $\cO_6$. 
Also, in the baryonic case only the transverse $\gamma^*$ polarisation contributes at leading power, while in the mesonic case the dominant annihilation effect appears for longitudinal $\gamma^*$ polarisation.

\section{Numerical results}
\setcounter{equation}{0}
\label{sec:num}

\begin{table}[t!]
\renewcommand{\arraystretch}{1.1}
\centering
\begin{tabular}{ccc}  
\toprule
\quad Parameter \quad & Value & \quad Ref. \quad \\
\midrule
$f_\Lambda$ & $(5.96_{-0.19}^{+0.20})\cdot 10^{-3} \GeV^2$ & \cite{RQCD:2019hps}  \\
$f_{\Lambda_b}^{(2)}$ & $(3.0 \pm 0.5) \cdot 10^{-2}  \GeV^3$ & \cite{Groote:1997yr,Ball:2008fw} \\
$s_0$ & $2.55 \GeV^2$ & \cite{Feldmann:2011xf} \\
$\omega_M (n_+\cdot p')$ & $2.5\pm 0.5 \GeV^2$ & \cite{Feldmann:2011xf} \\
$1/\omega_0$ & $3.4\pm 1.6 \MeV^{-1}$ & \cite{Feldmann:2011xf,Wang:2015ndk} \\
\bottomrule
\end{tabular}
\caption{Input parameters used to evaluate the LCSRs.
The lattice QCD calculation of $f_\Lambda$ in Ref.~\cite{RQCD:2019hps} is compatible with an older sum-rule estimate in Ref.~\cite{Liu:2008yg}. Notice that the normalisation convention in Ref.~\cite{RQCD:2019hps} differs from that used by us and in Ref.~\cite{Liu:2008yg} by a $SU(3)$ Clebsch-Gordan factor $\sqrt{\frac32}$.
}
\label{tab:inputs}
\end{table}

We provide numerical results for the non-local form factors $\H_{\perp(5)}^{(i)}$ and $\H_{+(5)}^{(i)}$ using the LCSRs in Eqs.~\eqref{eq:LCSRH5perp} and \eqref{eq:LCSRH5plus} and also taking into account the identities \eqref{eq:relH} (see also \refapp{match} for the integrated LCSRs).
These LCSRs are evaluated with the inputs listed in \reftab{inputs} and the LCDAs models obtained in \refapp{LCDAmodel}.
As there is no independent estimate of the LCDA parameter $\omega_0$, we vary its inverse in the interval $[1.8,5.0]\MeV^{-1}$.
This very conservative interval contains with margin the estimates of Refs.~\cite{Feldmann:2011xf,Wang:2015ndk}.
The interval for the Borel parameter $\omega_M$ is chosen in such a way that this parameter is both sufficiently large to suppress higher power corrections in the OPE and sufficiently small to ensure that the contribution of the continuum and excited states is subleading compared to that of the $\Lambda$ baryon.
We use the same central value of Ref.~\cite{Feldmann:2011xf} and vary it within $\pm 0.5 \GeV^2$.
We have checked that our LCSRs are stable in this interval for the Borel parameter.
As in Ref.~\cite{Feldmann:2011xf}, we choose $s_0$ to be equal to the mass of the next resonance with the same quantum numbers of the $\Lambda$ baryon.
All parameters in \reftab{inputs} are assumed to be Gaussian distributed, except for the Borel parameter for which we take a flat distribution.

The LCSRs in Eqs.~\eqref{eq:LCSRH5perp} and \eqref{eq:LCSRH5plus} can be used for values of $q^2$ such that the energy of the $\Lambda$ baryon is of the order of $m_{\Lambda_b}/2$ in the $\Lambda_b$ baryon rest frame.
To avoid large violations of the quark-hadron duality, we also take $q^2$ larger than the narrow vector resonances such as the $\rho$ and $\omega$ mesons.
We can therefore evaluate our LCSRs in the range $2 \GeV^2 \lesssim q^2\lesssim 6 \GeV^2$. 
We choose the following $q^2$ points: $q^2 = \{2,4,6\}\GeV^2$. 
We obtain 
\begin{equation}
\begin{aligned}
    \label{eq:resH}
    &
    10^{5}\cdot \H_\perp^{(5)} 
    (2 \GeV^2)
    = -(1.2 \pm 5.6) - i (15.6 \pm 8.2)
    ,
    &
    \\
    &
    10^{5}\cdot \H_\perp^{(5)} 
    (4 \GeV^2)
    = (6.8 \pm 5.3) - i (8.2 \pm 2.6)
    ,
    &
    \\
    &
    10^{5}\cdot \H_\perp^{(5)} 
    (6 \GeV^2)
    = (6.5 \pm 3.2) - i (3.9 \pm 1.6)
    ,
    &
\end{aligned}
\end{equation}
and
\begin{equation} 
\begin{aligned}
    &
    10^{7}\cdot \H_+^{(5)} 
    (2 \GeV^2) =
    -(1.31 \pm 0. 74) + i (1.47 \pm 0.38) 
    ,
    &
    \\
    &
    10^{7}\cdot \H_+^{(5)} 
    (4 \GeV^2) =
    -(2.48 \pm 0. 88) + i (1.16 \pm 0.45) 
    ,
    &
    \\
    &
    10^{7}\cdot \H_+^{(5)} 
    (6 \GeV^2) =
    -(3.38 \pm 0. 90) + i (0.85 \pm 0.58) 
    ,
    &
\end{aligned}
\end{equation}
We remind the reader that the results for the other non-local form factors can be obtained using the identities \eqref{eq:relH}.
We can cast the results above also in the form of a $q^2$-dependent shift to $C_9$ using \refeqs{deltaC9a}{deltaC9d}:
\begin{equation}
\begin{aligned}    
    \label{eq:resDel}
    &
    10^{2}\cdot \Delta C_{9,\perp}
    (2 \GeV^2)
    = (0.6 \pm 2.5) + i (6.9 \pm 1.3)
    ,
    &
    \\
    &
    10^{2}\cdot \Delta C_{9,\perp}
    (4 \GeV^2)
    = -(0.97 \pm 0.98) + i (1.89 \pm 0.62)
    ,
    &
    \\
    &
    10^{2}\cdot \Delta C_{9,\perp}
    (6 \GeV^2)
    = -(0.64 \pm 0.39) + i (0.63 \pm 0.42)
    ,
    &
\end{aligned}
\end{equation} 
and
\begin{equation}
\begin{aligned}
    &
    10^{5}\cdot \Delta C_{9,+}
    (2 \GeV^2) =
    (5.2 \pm 3.1) - i (8.7 \pm 3.8) 
    ,
    &
    \\
    &
    10^{5}\cdot \Delta C_{9,+}
    (4 \GeV^2) =
    (5.1 \pm 2.2) - i (4.1 \pm 4.0) 
    ,
    &
    \\
    &
    10^{5}\cdot \Delta C_{9,+}
    (6 \GeV^2) =
    (4.3 \pm 2.1) - i (2.3 \pm 3.7) 
    .
    &
\end{aligned}
\end{equation}
For the local form factors we have used the analytical results of Ref.~\cite{Feldmann:2011xf}, i.e. \refeqa{xirel}{xiexpl}, evaluated with the inputs of \reftab{inputs}.
The values of the Wilson coefficients --- evaluated at the scale $\mu = m_b$ --- are taken from Ref.~\cite{Buchalla:1995vs}.

\noindent A few comments on our numerical results are in order:
\begin{itemize}
    \item 
    As expected from the OPE results obtained in \refsec{finalOPE}, $\H_+^{(5)}$ is power suppressed w.r.t. $\H_\perp^{(5)}$ and this is reflected in the numerical results of \refeq{resH}.
    This makes the contribution of $\H_+^{(5)}$ essentially negligible.
    \item 
    The uncertainty of our numerical results is dominated by the  LCDA model parameter $\omega_0$.
    It is therefore crucial to have a better knowledge of the LCDAs and their parameters in order to improve the accuracy of the current calculation.
    \item 
    We find that $|\Delta C_{9,\perp}|/C_9$ is $\order{1\%}$. 
    This means that if, as expected, the theoretical precision of the local $\Lambda_b\to \Lambda$ form factors is improved by future lattice QCD calculations~\cite{Meinel:2023wyg}, the annihilation topologies should be taken into account in the prediction of $\Lambda_b\to \Lambda \ell^+ \ell^-$ observables.
    \item 
    Comparing the $C_9$ shift due to the weak annihilation in $B^+\to K^{(*)+}\ell^+\ell^-$ decays~\cite{Bobeth:2007dw,Beneke:2001at,Khodjamirian:2012rm}, with our $\Delta C_{9,\perp}$, we find that these two different contributions have a very similar $q^2$ dependence.
    In particular, both their real parts have a zero at $q^2\simeq 2\GeV^2$, while the imaginary parts are positive definite.
    Regarding their magnitude, we find $\Delta C_{9,\perp}$ to be about 5 times larger than for $B^+\to K^{+}\ell^+\ell^-$  and $B^+\to K_\parallel^{*+}\ell^+\ell^-$ decays.
\end{itemize}

\newpage

\section{Conclusions}
\label{sec:conc}

In this work, we have studied the non-local contributions of the strong penguin operators  
in $\Lambda_b \to \Lambda \ell^+\ell^+$ decays, 
where a virtual photon is radiated from one of the light quarks.
We refer to this situation as ``annihilation topologies'' because of the analogy with annihilation in $B \to K^{(*)} \ell^+\ell^+$ decays. 
Their contribution to the corresponding non-local form factors is calculated using light-cone sum rules (LCSRs) with $\Lambda_b$ light-cone distribution amplitudes (LCDAs).
More precisely, we find that --- at leading power --- the hard-scattering kernel entering the factorisation formula for the underlying correlator
depends on opposite light-cone projections of the two light-quark momenta in the $\Lambda_b$ baryon.
This implies that in this case the required hadronic information about the $\Lambda_b$ bound state is contained in a new type of soft functions that generalise  the standard LCDAs which are known in the literature and used, for instance, in local form-factor calculations.
In order to evaluate the LCSRs, we have constructed a model for these new soft functions that links them to the standard $\Lambda_b$ LCDAs.
On this basis, we have presented the result from the leading-order LCSRs for the annihilation contribution to the non-local form factors in analytical form.
Here we have focused on the leading annihilation topology, where the virtual photon is radiated from one of the light (soft) quarks in the 
$\Lambda_b$ baryon.
Numerical predictions are presented in the form of a $q^2$-dependent shift $\Delta C_9$ of the Wilson coefficient $C_9$, where $q^2$ is the invariant mass of the lepton pair.
In the considered range $q^2 \in [2,6]$, we observe that $|\Delta C_{9,\perp}|/C_9 \sim \order{1\%}$ and ${\rm Im} \Delta C_9 \neq 0$, 
and hence this effect should not be neglected in precision analyses of $\Lambda_b \to \Lambda \ell^+\ell^+$ observables.

Our findings show a number of analogies with the annihilation topologies in $B \to K^{(*)} \ell^+\ell^+$ decays.
For instance, in both cases $\Delta C_9$ features a very similar $q^2$ dependence and is of similar numerical size, which can be traced back
to the functional form of the intermediate hard-collinear light-quark propagator folded with the modelled shape of the (generalised) LCDAs.
A major difference is that in the mesonic case the operators of the weak effective Hamiltonian that contribute at leading order are $\cO_3$ and $\cO_4$ (with left-handed $q\bar q$ currents), 
while in the baryonic case they are $\cO_5$ and $\cO_6$ (with right-handed $q\bar q$ currents).
This can be traced back to the Dirac structure of the penguin operators that results from replacing the light quark fields by their leading spinor components in soft-collinear effective theory.

We re-emphasise that in our analysis we only considered one particular non-factorising decay topology, and it is left for future work to perform similar investigations for the sub-leading annihilation topologies, but also to study related topologies where a quark loop originating from the 4-quark operators or the chromomagnetic penguin operator connects to one of the light quarks from the $\Lambda_b$ bound state by hard-collinear gluon exchange (similar topologies had been studied for mesonic transition in the past). While in both cases, we expect sub-leading numerical effects, verification by explicit calculation would be desirable.

To conclude,  the inclusion of genuinely \emph{non-factorising} contributions from hadronic operators 
in $\Lambda_b \to \Lambda \ell^+\ell^-$ decays at large recoil are important, not only 
to improve the accuracy of SM predictions for physical observables and to sharpen the current constraints on physics beyond the SM,
but also  as a laboratory to test our understanding and further deepen our knowledge of non-perturbative QCD effects in exclusive baryonic reactions.

\section*{Acknowledgements}

We thank Marzia Bordone for her contributions in the early stage of this project.
This research is supported by the Deutsche Forschungsgemeinschaft (DFG, German Research Foundation) under grant 396021762 -- TRR 257.
The work of N.G. has been partially supported by STFC consolidated grants ST/T000694/1 and ST/X000664/1.

\appendix 
\addtocontents{toc}{\protect\setcounter{tocdepth}{1}}

\renewcommand{\theequation}{\thesection.\arabic{equation}}

\appendix

\section[Models for \texorpdfstring{$\boldsymbol{\Lambda_b}$}{} LCDAs in momentum space]{Models for \texorpdfstring{$\boldsymbol{\Lambda_b}$}{} LCDAs in momentum space}
\setcounter{equation}{0}
\label{app:LCDAmodel}

A straightforward procedure for constructing explicit models for the LCDAs appearing in the momentum-space projector \eqref{eq:newlcproj} has been outlined in Ref.~\cite{Bell:2013tfa}. 
For this purpose, the following ansatz is used
\begin{align}
M^{(2)}(v,k_1,k_2) = \tilde \psi_v(x_1,x_2, K^2) \, \slashed k_2 \, \slashed v \, \slashed k_1 
\,,
\label{eq:lcproj}
\end{align}
where
\begin{align}
    & x_i = 2 \, v\cdot k_i \,, \qquad K^2 = (k_1+k_2)^2 \,.
\end{align}
In this way the Dirac matrix $M^{(2)}$ automatically fulfils the equations of motion for on-shell light quarks in the $\Lambda_b$ Fock state.\footnote{The resulting relations between the individual LCDAs/soft functions are sometimes referred to as ``Wandzura-Wilczek approximation''. These are valid up to corrections involving the four-particle LCDAs.}
We further follow Ref.~\cite{Bell:2013tfa} and \emph{assume} that the shape of the wave function $\psi_v$
is predominantly determined by its dependence on the invariant mass of the three-quark bound state, $(m_b v + k_1 +k_2)^2 \simeq m_b^2 + m_b \, (x_1+x_2)$, such that the $K^2$-dependence can be ignored,
\begin{align}
    \psi_v(x_1,x_2, K^2) \to  \psi_v(x_1,x_2) \,.
\end{align}
We may then fold the ansatz for the momentum-space projector in (\ref{eq:lcproj})
with a test kernel which includes terms at most linear in the transverse momenta $k_{i\perp}$.
Writing 
\begin{align}
 \slashed k_1 & = \bar\omega_1 \, \frac{\slashed n_-}{2} + (x_1 - \bar\omega_1) \, \frac{\slashed n_+}{2} + \slashed k^\perp_{1} \,, 
 \cr 
    \slashed k_2 & = \omega_2 \, \frac{\slashed n_+}{2} + (x_2 - \omega_2) \, \frac{\slashed n_-}{2} + \slashed k^\perp_{2} \,,
\end{align}
with $k_1^2=k_2^2=0$,
this leads to
\begin{align}
 & \int \widetilde{dk_1} \, \int \widetilde{dk_2} \,
{\rm tr}\left[\left( T_0(\bar{\omega}_1,\omega_2) + k_{i\perp}^\mu T_\mu^i(\bar{\omega}_1,\omega_2) \right)
 M^{(2)}(v,k_1,k_2) \right]
 \cr = &  \int d\bar{\omega}_1 \, d\omega_2 \, \int\limits_{\bar{\omega}_1}^\infty dx_1 \int\limits_{\omega_2}^\infty dx_2  \Bigg\{
\cr
& \quad
\, {\rm tr} 
\left[T_0(\bar{\omega}_1,\omega_2) 
\left(
    (x_1 - \bar{\omega}_1) \omega_2 \, \frac{\slashed n_+}{2} + \bar{\omega}_1 (x_2-\omega_2) \, \frac{\slashed n_-}{2} 
\right)
\right]  
\cr
& -
{\rm tr} \left[ T_\mu^1(\bar{\omega}_1,\omega_2)
\left( 
    \bar{\omega}_1 \omega_2  (x_1-\bar{\omega}_1) \, \frac{\slashed n_+\slashed n_-}{4}
    + \bar{\omega}_1 (x_1-\bar{\omega}_1)(x_2-\omega_2) \, \frac{\slashed n_-\slashed n_+}{4} 
\right) \frac{\gamma_\perp^\mu}{2}
\right]  
\cr
& -
{\rm tr} 
\left[ T_\mu^2(\bar{\omega}_1,\omega_2) \,\frac{\gamma_\perp^\mu}{2} 
\left( 
    \omega_2 (x_1-\bar{\omega}_1) (x_2-\omega_2) \, \frac{\slashed n_-\slashed n_+}{4}
    + \bar{\omega}_1 \omega_2 (x_2-\omega_2) \, \frac{\slashed n_+\slashed n_-}{4} 
\right)
\right]  
\cr
& \qquad \Bigg\} \, \psi_v (x_1,x_2)
 \,,
 \label{eq:proj2}
\end{align}
where $\widetilde{dk}_i = d^3k_i/(\pi x_i)$ are  Lorentz-invariant phase-space integrals for the
(massless) light quarks in the $\Lambda_b$ baryon.
Comparing with the momentum-space projector (\ref{eq:newlcproj}), one easily obtains
\begin{align}
\chi_2(\bar{\omega}_1,\omega_2)
&=
\int\limits_{\bar{\omega}_1}^\infty dx_1 \int\limits_{\omega_2}^\infty dx_2 \,
\left(
x_1 \omega_2 + x_2 \bar\omega_1 - \bar\omega_1 \omega_2 - \frac{x_1 x_2}{2}
\right)
\psi_v(x_1,x_2)
\,,
\end{align} 
and
\begin{align}
 \chi_{42}^{(i)}(\bar{\omega}_1,\omega_2)&= \frac12 \, \int\limits_{\bar{\omega}_1}^\infty dx_1 \int\limits_{\omega_2}^\infty dx_2 \,
   x_2 \, (x_1-2\bar{\omega}_1) \, \psi_v(x_1,x_2) \,,
   \\
 \chi_{42}^{(ii)}(\bar{\omega}_1,\omega_2)&= \frac12 \, \int\limits_{\bar{\omega}_1}^\infty dx_1 \int\limits_{\omega_2}^\infty dx_2 \,
   x_1 \, (x_2-2\omega_2) \, \psi_v(x_1,x_2) \,,
    \\
  \chi_X(\bar{\omega}_1,\omega_2)&= \frac12 \, \int\limits_{\bar{\omega}_1}^\infty dx_1 \int\limits_{\omega_2}^\infty dx_2 \,
    (x_1-2\bar{\omega}_1)\, (x_2-2\omega_2) \, \psi_v(x_1,x_2) \,.
\end{align}

A simplified model for the baryon wave functions can then be obtained by assuming an exponential
dependence of the wave function $\psi_v$ on $(x_1 + x_2)$ as in Refs.~\cite{Feldmann:2011xf,Bell:2013tfa}:
\begin{align}
    \psi_v(x_1,x_2)
    :=
    \frac{1}{\omega_0}
    \exp\left(-\frac{x_1+x_2}{\omega_0^6} \right)
    \,.
\end{align}
For this case, we obtain the following expressions for the linear combinations entering the momentum-space projector, 
\begin{align}
\chi_2(\bar\omega_1,\omega_2) + \chi_{42}^{(i)}(\bar\omega_1,\omega_2) &= \frac{\omega_2}{\omega_0^3} \, 
e^{-(\bar \omega_1+\omega_2)/\omega_0} \,, \\
\chi_2(\bar\omega_1,\omega_2) + \chi_{42}^{(ii)}(\bar\omega_1,\omega_2) &= \frac{\bar \omega_1}{\omega_0^3} \, 
e^{-(\bar \omega_1+\omega_2)/\omega_0} \,, \\
\bar\chi_{42}^{(i)}(\bar\omega_1,\omega_2) - \bar\chi_X(\bar\omega_1,\omega_2) &= \frac{\bar \omega_1 \omega_2}{\omega_0^3} \, 
e^{-(\bar \omega_1+\omega_2)/\omega_0}  \,, \\
\bar\chi_{42}^{(i)}(\bar\omega_1,\omega_2) + \bar\chi_X(\bar\omega_1,\omega_2) &= \frac{\bar \omega_1}{\omega_0^2} \, 
e^{-(\bar \omega_1+\omega_2)/\omega_0}  \,, \\
\hat\chi_{42}^{(ii)}(\bar\omega_1,\omega_2) + \hat\chi_X(\bar\omega_1,\omega_2) &= \frac{\omega_2}{\omega_0^2} \, 
e^{-(\bar \omega_1+\omega_2)/\omega_0}   \,, \\
\hat\chi_{42}^{(ii)}(\bar\omega_1,\omega_2) - \hat\chi_X(\bar\omega_1,\omega_2) &= \frac{\bar\omega_1\omega_2}{\omega_0^3} \, 
e^{-(\bar \omega_1+\omega_2)/\omega_0}  \,. 
\end{align} 

\section{Further details on the LCSRs}
\setcounter{equation}{0}
\label{app:match}

In the following, we provide a few intermediate steps in the derivation of the LCSRs in \refsec{LCSR}, i.e. the matching of the OPE calculation $\Pi_\mu^\OPE$ of \refeqa{LPImperp}{LPIpmu} onto the respective hadronic representations of \refeqa{hadrep-p}{hadrep-g}.
This matching yields the LCSRs for the non-local form factors $\H_{\perp(5)}^{(i)}$ and $\H_{+(5)}^{(i)}$.
After applying quark-hadron duality and performing the Borel transform, the resulting LCSRs read
\begin{align}
    &
    M_{\Lambda_b}^2
    f_\Lambda (n_+\cdot p' - m_\Lambda)
    \frac{\slashed{n}_-}{2}
    \left[
    \H_\perp^{(i)}(q^2) 
     \, \gamma_\mu^\perp 
    -  \H_{\perp5}^{(i)}(q^2) 
    \gamma_\mu^\perp  
    \gamma_5 
    \right] 
    e^{-\frac{m_\Lambda^2}{(n_+\cdot p') \, \omega_M}}
    u_{\Lambda_b}(p)
    =
    \nonumber\\
    &
\ara{\mp} \frac{(Q_u + Q_d) f_{\Lambda_b}^{(2)} }{192\pi^2} \,
    \int\limits_0^{\sigma_0} d\sigma \int\limits_0^\sigma d\omega_2
    \int\limits_0^\infty d\bar\omega_1 \, 
    \frac{(n_+ \cdot p')^2}{n_+ \cdot q - \bar \omega_1 + i\epsilon} 
  \left( \chi_2(\bar \omega_1,\omega_2) + \chi_{42}^{(ii)}(\bar\omega_1,\omega_2) \right) 
    \cr & 
    \times 
      \left( - \frac{g_{\mu\nu}^\perp}{2} \mp \frac{i \epsilon_{\mu\nu\rho\sigma} \, n_-^\rho n_+^\sigma}{4} \right)
    e^{-\frac{\sigma}{\omega_M} }
      \left[
    \slashed n_- 
    \gamma^\nu_\perp P_L
      \right] u_{\Lambda_b}(v)
      \,,
\end{align}
and
\begin{align}
    & - 
    \frac{M_{\Lambda_b}^2}{2 q^2}
    f_\Lambda (n_+\cdot p' + m_\Lambda)
    \cr
    & \times
    \frac{\slashed{n}_-}{2}
    \left[
        s_- (M_{\Lambda_b}+ m_\Lambda)\,\H_+^{(i)}(q^2)
        - s_+ (M_{\Lambda_b} - m_{\Lambda})
        \H_{+5}(q^2) \gamma_5
    \right]
    e^{-\frac{m_\Lambda^2}{(n_+\cdot p') \, \omega_M}}
    u_{\Lambda_b}(p)
    \cr
    & =
    \ara{\pm} \frac{1}{4} \frac{(Q_u + Q_d)  f_{\Lambda_b}^{(2)} }{192\pi} \, M_{\Lambda_b} 
    \int\limits_0^{\sigma_0} d\sigma \int\limits_0^\sigma d\omega_2
    \int\limits_0^\infty d\bar\omega_1 \, 
    \frac{n_+\cdot p'}{n_+ \cdot q - \bar \omega_1 + i\epsilon} 
    \left( 
        \hat\chi_{42}^{(ii)}(\bar{\omega}_1,\omega_2) + \hat\chi_X(\bar{\omega}_1,\omega_2) 
    \right)
    \nonumber \\*
    & \times 
    \left[\gamma_\beta \slashed n_- \right]\left( \frac{g^{\beta\nu}_\perp}{2} \pm \frac{i \epsilon^{\beta\nu\{ n_+\}\{n_-\}}}{4} \right)  \left[
        \slashed n_+
        \gamma_\nu P_L
    \right]  
    e^{-\frac{\sigma}{\omega_M} } u_{\Lambda_b}(v)
    \,.
\end{align}
Here $\sigma_{(0)} \equiv s_{(0)}/(n_+ \cdot p')$ and we have also added the contribution of the diagram where the photon is emitted from the $d$ quark instead of the $u$ quark, resulting in the total charge factor $(Q_u+Q_d)$.
It is a straightforward task to derive the identities~\eqref{eq:relH} from these LCSRs using Eqs.~\eqref{eq:gid1}, \eqref{eq:gid2}, \eqref{eq:gid3}, and \eqref{eq:gid4}. 

Using the model for the LCDAs presented in \refapp{LCDAmodel}, it is possible to perform all integrations over light-cone momenta analytically. 
Here, one has to take into account that, the integration over $d\bar{\omega}_1$ contains a singularity on the integration path for $\epsilon\to 0$.
These integrals can be performed by Cauchy's theorem, along the same lines as for the annihilation in $B\to K^{(*)}\ell^+\ell^-$ \cite{Beneke:2001at}, leading to
\begin{multline}
    \label{eq:LCSRH5perpexpl}
    \H_\perp^{(5)}(q^2) 
    = 
    \frac{f_{\Lambda_b}^{(2)}}{f_\Lambda} 
    \frac{(Q_u + Q_d) }{192\pi^2}
    \frac{\omega_M (n_+\cdot p')^2}{ M_{\Lambda_b}^2 (n_+\cdot p' - m_\Lambda)}
    \frac{
        \omega_0 \big(e^{\frac{\sigma_0}{\omega_0}} -1 \big)
        - \omega_M e^{\frac{\sigma_0}{\omega_0}} \big(e^{\frac{\sigma_0}{\omega_M}}-1\big)
    }
    {\omega_0^2 (\omega_0 + \omega_M)}
    \\
    \times
    \left(
        \omega_0 e^{\frac{M_{\Lambda_b}}{\omega_0}} + (n_+\cdot p' - M_{\Lambda_b})
        e^{\frac{n_+\cdot p'}{\omega_0}}
        \left(
            {\rm Ei} \left(\frac{M_{\Lambda_b} - n_+\cdot p'}{\omega_0}\right) -i\pi
        \right)
    \right)
    e^{
        \frac{m_\Lambda^2 - \sigma_0\, n_+\cdot p' }{\omega_M \, n_+\cdot p'}
        - \frac{M_{\Lambda_b} + \sigma_0}{\omega_0}
    }
    \,,
\end{multline} 
and
\begin{multline} 
    \H_+^{(5)}(q^2)
     =
    \frac{f_{\Lambda_b}^{(2)}}{f_\Lambda} 
    \frac{(Q_u + Q_d) }{192\pi^2}
    \frac{\omega_M \, n_+\cdot p'}{ s_- ( m_\Lambda + M_{\Lambda_b}) }
    \frac{
        n_+\cdot p' - M_{\Lambda_b}
    }
    {m_\Lambda + n_+\cdot p'}
    \frac{
            {\rm Ei} \left(\frac{M_{\Lambda_b} - n_+\cdot p'}{\omega_0}\right) -i\pi
    }{(\omega_0 + \omega_M)^2}
    \\
    \times
    \left(
        (\omega_0^2 + 2\omega_0\omega_M) \left(1 - e^{\frac{\sigma_0}{\omega_0}} \right)
        - \omega_M^2   e^{\frac{\sigma_0}{\omega_0}} \left(1 - e^{\frac{\sigma_0}{\omega_M}} \right)
        + \sigma_0 (\omega_0 + \omega_M)
    \right)
    e^{
        \frac{m_\Lambda^2 - \sigma_0\, n_+\cdot p' }{\omega_M \, n_+\cdot p'}
        - \frac{M_{\Lambda_b} - n_+\cdot p' + \sigma_0}{\omega_0}
    }
    \,,
    \label{eq:LCSRH5plusexpl}
\end{multline}
where ${\rm Ei}(x)$ is the exponential integral function.

\bibliographystyle{JHEP}
\bibliography{references}
 
\end{document}